\documentclass[12pt]{iopart}
\usepackage{graphicx}
\usepackage{amssymb}
\usepackage[english]{babel}
\usepackage{psfrag}
\newcommand{\bq}{\begin{equation}}
\newcommand{\eq}{\end{equation}}
\newcommand{\ba}{\begin{eqnarray}}
\newcommand{\ea}{\end{eqnarray}}

\usepackage{ifthen}

\newcommand{\pd}[2]{
\ifthenelse{\equal{#2}{1}}{\frac{\partial}{\partial #1}}
{\frac{\partial ^ #2}{\partial #1 ^#2}}
}
\newcommand{\dd}[2]{
\ifthenelse{\equal{#2}{1}}{\frac{{\rm d}}{{\rm d} #1}}
{\frac{{\rm d}^ #2}{{\rm d} #1 ^#2}}
}

\newcommand{\pf}[3]{
\ifthenelse{\equal{#3}{1}}{\frac{\partial #1 }{\partial #2}}
{\frac{\partial ^ #3 #1 }{\partial #2 ^#3}}
}
\newcommand{\df}[3]{
\ifthenelse{\equal{#3}{1}}{\frac{{\rm d} #1}{{\rm d} #2}}
{\frac{{\rm d} ^ #3 #1}{{\rm d} #2 ^#3}}
}

\begin{document}
\title{Condensation in stochastic mass transport models: beyond the 
zero-range process}
\author{M. R. Evans and B. Waclaw}
\address{SUPA, School of Physics and Astronomy, University of Edinburgh, 
Mayfield Road, Edinburgh EH9 3JZ, United Kingdom}
\eads{\mailto{mevans@staffmail.ed.ac.uk}, \mailto{bwaclaw@staffmail.ed.ac.uk}}

\begin{abstract}
We consider an extension of the zero-range process to the case where the hop 
rate depends on the state of both departure and arrival sites. We recover 
the misanthrope and the target process as special cases for which the 
probability of the steady state factorizes over sites.
We discuss conditions which lead to the condensation of particles and show 
that although two different hop rates can lead to the same steady state,
they do so with  sharply contrasting  dynamics. The first case resembles the 
dynamics of the zero-range process, whereas the second case, in which the 
hop rate increases with the occupation number of both sites, is similar to 
instantaneous gelation models. This new ``explosive'' condensation reveals 
surprisingly rich behaviour, in which the process of condensate's formation 
goes through a series of collisions between clusters of particles moving 
through the system at increasing speed. We perform a detailed numerical and 
analytical study of the dynamics of condensation: we find the speed of the 
moving clusters, their scattering amplitude, and their growth time. We 
finally show that the time to reach steady state decreases with the size of 
the system.
\end{abstract}
Date \today
\maketitle

\section{Introduction}

Condensation is  one of the most ubiquitous forms of phase transition 
experienced in everyday life. Water vapour condenses into a liquid on cold 
surfaces such as windows or a cold beverage taken out of the fridge. Fog is an 
aerosol that occurs when temperature drops below the dew point and water 
vapour condenses into tiny droplets suspended in air. Condensation can also 
be seen as small ``clouds'' in low-pressure zones above aircraft wings when 
flying in damp weather conditions. Condensation is also extensively used in 
industry for making liquid oxygen and nitrogen, production of liquefied 
natural gas, or in petroleum distillation.

All these examples describe a transition from a gas to a liquid state of 
matter. However, in statistical physics, the notion of condensation is  more 
widely used to describe any process in which a finite fraction of some 
conserved quantity becomes localized in the phase space. So understood, 
condensation can occur either in real space or in momentum space. In fact, 
the most prominent example of statistical-physics condensation - 
Bose-Einstein condensation - takes place in the momentum space where a 
finite fraction of all particles present in the system assumes the lowest 
energy state. Such generalized condensation occurs in many different 
systems, both in- and out-of-equilibrium: granular clustering, wealth 
condensation \cite{BJJKNPZ02}, quantum gravity \cite{bbpt}, polydisperse 
hard spheres \cite{EMPT10}, hub formation in complex networks \cite{redner}, 
or even jamming in traffic flow \cite{OEC,CSS}.

Perhaps surprisingly, many features of the generalized condensation 
transitions observed  in the  systems listed in the previous paragraph, can 
be captured by a very simple model called the zero-range process (ZRP) 
\cite{Spitzer}. ZRP is a well-studied, non-equilibrium model of particles 
hopping between sites of a lattice. In the simplest  one-dimensional 
asymmetric version of this model, a particle moves from site $i$ to $i+1$ of 
a one-dimensional periodic lattice with rate
$u(m_i)$, where $m_i$ is the occupancy of the departure site $i$.
As the rates are totally asymmetric a current always flows  and detailed 
balance cannot be satisfied,
thus the steady state is  non-equilibrium in nature. When the hop rate 
$u(m)$ decays slowly enough with $m$ and the density of particles is larger 
than a certain critical density (the exact value of which  depends on the 
function $u(m)$), a finite fraction of all particles condenses onto a 
single, randomly chosen site.

The advantage of ZRP is that its non-equilibrium steady state has a simple 
factorised form which is amenable to analysis and has furnished our
understanding of the condensation transition. This is why the model is often 
used as an effective description of more complicated systems, for example
systems with more microscopic degrees of freedom, where such a complete 
analytical description is often impossible.

The characteristic feature of the ZRP is that the hop rate  depends only on
the departure site and not on the destination site.
A natural generalisation is to consider a hop rate
$u(m,n)$ where $m$ is the occupancy of the departure site $i$ and
$n$ is the occupancy of the destination site $i+1$. Such processes are 
sometimes referred to as migration processes in the mathematical literature 
\cite{fpkelly}. An example of such process is the `misanthrope process' 
\cite{CT85} whose name originally derived from considering  a hop rate $u(m,n)$ as an increasing function of $m$ and decreasing function $n$; an effective repulsion between individuals results. 
The name  `misanthrope process' is now generally used for any hop rate
of the form $u(m,n)$.
Such a  process has been used, for 
example, to model link rewiring in complex network models \cite{EH05}. 
Another example  is the target process introduced in 
Ref. \cite{godr-luck}. Here, the rate $u(m,n) = (1-\delta_{m,0})v(n)$, with 
some function $v(n)>0$, depends on the occupation of the arrival site.

\begin{figure}
\center
\psfrag{m}{$m$} \psfrag{n}{$n$} \psfrag{umn}{$u(m,n)$}
\includegraphics*[width=10.0cm]{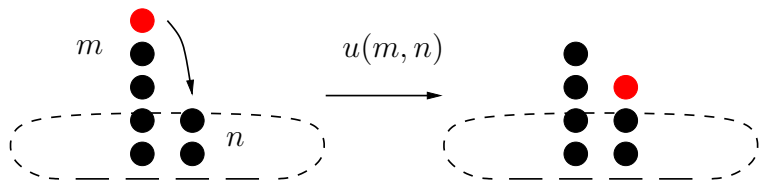}
\caption{Definition of the misanthrope process: a particle hops from site 
with $m$ to site with $n$ particles with rate $u(m,n)$.
\label{model_def}}
\end{figure}

Our aim in this paper is to analyse a more general case of $u(m,n)$. It has 
been shown that the misanthrope process has a factorised steady state when 
certain conditions on $u(m,n)$ (expressed below in Eqs.~(\ref{c1}) and 
(\ref{c2})) are satisfied. Here we provide examples of hop rates when these 
conditions are satisfied. A particularly interesting case that we discuss is 
when the hop rate factorizes:
\bq
  u(m,n) = w(m) v(n). \label{facu}
\eq
We find that for (\ref{facu}) to yield a factorised steady state requires a 
further condition (\ref{ufact}) on the functions $w(m)$ and $v(n)$.

We also analyse the conditions for condensation to occur. As the steady 
state has a factorised form  the conditions on the weights $f(m)$ will be 
the same as for the zero-range process. However, there is more freedom in 
the misanthrope dynamical rules with which to realise this asymptotic 
behaviour. Moreover, the existence of condensation depends not only on the 
asymptotic behaviour of $u(m,n)$ as it does for the ZRP, but also on 
$u(m,0)$ which gives the hop rate to an empty site. In the case of 
factorized hop rate (\ref{facu}), the condensation criterion turns out to 
depend on $v(0)$.

We also expand on our previous work \cite{prl} and study the dynamics of 
condensation for two different choices of the factorized hop rate 
(\ref{facu}). Both forms of $u(m,n)$ lead to the same steady-state 
probability distribution but contrasting dynamical behaviour. A particularly 
interesting case, in which $u(m,n)$ increases with both $m$ and $n$, leads 
to ``explosive condensation'' where the time to condensation {\it decreases} 
with increasing system size. This is very similar to ``instantaneous 
gelation'' known from the theory of coagulation processes, and we discuss 
similarities and differences between the two models in the last section.

Beyond factorised steady states it is known that pair-factorised states 
(also referred to as Markov measures) may occur in some stochastic models.
A simple example is the 1+1d solid-on-solid model \cite{sos,sos2,sos3} with a ``pinning potential'', 
which describes the height of the interface $\{m_i\}$ between two phases and whose microstate probability reads
\bq
	P(m_1,\dots,m_N) = \exp\left( -J\sum_i |m_i-m_{i+1}| + U \sum_i \delta_{m_i,0}\right), \label{eq:sos}
\eq
with two parameters $J,U$. The $J$-term describes the tendency of the system to form straight interface (``surface tension'') whereas $U$ gives the energy cost for the interface having zero height. The model can be used for example to describe the formation of a new layer of atoms on a solid substrate. The probability (\ref{eq:sos}) can be recast into
\bq
P(\{m_i\})= \prod_{i=1}^L g(m_i,m_{i+1}), 
\label{pairfact}
\eq
where $g(m,n)=\exp\left[ -J |m-n| + (U/2) (\delta_{m,0} + \delta_{n,0})\right]$ is the pairwise weight.
Although this model is an equilibrium model, more recently a class of nonequilibrium models which exhibit such steady 
states have also been proposed \cite{EHM06}. 
One could hope that    if conditions (\ref{c1}),  (\ref{c2}) are not 
satisfied
there might be some choices of  rates $u(m,n)$  which yield pair-factorised 
states.
We shall show that actually this is not the case and that either (\ref{c1}), 
(\ref{c2}) is satisfied
or the steady state has an unknown form.

\section{Misanthrope process\label{mis}}
We now define the model that we consider. $M$ particles reside on sites of a 
one-dimensional closed, periodic  chain of length
$L$. Each site $i$ carries $m_i$ particles, so that $\sum_{i=1}^L
m_i=M$. A particle hops from site $i$ to site $i+1$ with rate
$u(m_i,m_{i+1})$ which depends on the occupancies of both the departure
and the arrival site, see Fig.~\ref{model_def}. Periodicity implies
$m_{L+1}=m_1$.

\begin{figure}
\center
\psfrag{umn1}{$u(3,1)$} \psfrag{umn2}{$u(1,0)$} \psfrag{umn3}{$u(2,0)$}
\includegraphics*[width=10.0cm]{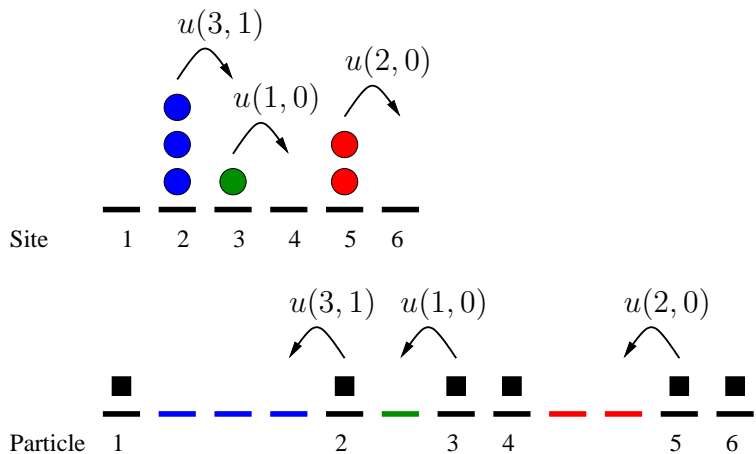}
\caption{Mapping between the misanthrope process and the total asymmetric 
simple exclusion process (TASEP).
\label{mapping_to_tasep}}
\end{figure}

The misanthrope process can be mapped to the total asymmetric simple 
exclusion process (TASEP) on a closed chain.
In Fig.~\ref{mapping_to_tasep} we show how this can be done. In TASEP, each 
site of a periodic 1d lattice is occupied by at most one particle. We can 
identify the number of particles in the misanthrope process with the number 
of vacancies between the particles in TASEP. To obtain the same dynamics as 
in the misanthrope process, we assume that the particle jumps to the left 
with rate $u(m,n)$ depending on the number of empty sites in front ($m$) and 
behind ($n$) the particle.

It is known that under certain conditions the steady state probability of 
the misanthrope process is given by a factorised form
\bq
P(\{m_i\})= \frac{1}{Z_L}\prod_{i=1}^L f(m_i)\, \delta_{M, \sum_j m_j}, \label{fullfact}
\eq
where $f(m)$ is a certain
non-negative weight function and $Z_L$ is the normalization
\bq
{Z_L}= \sum_{\{m_i=0\}}^{\infty} \prod_{i=1}^L f(m_i)\, \delta_{M, \sum_j m_j}. \label{norm}
\eq
 In (\ref{fullfact}) the Kronecker delta 
imposes the constraint that the total mass (number of particles) $M$
in the system is conserved. The conditions for factorisation can be derived 
as follows. In the steady state the equation
balancing probability currents from and to a given configuration is
\ba
\sum_{i=1}^L u(m_i,m_{i+1}) P(\dots,m_{i-1},m_i,\dots)= \nonumber \\ 
\sum_{i=1}^L
u(m_{i-1}+1,m_i-1) P(\dots,m_{i-1}+1,m_i-1,\dots), \label{pbalance}
\ea
where $P(\{m_i\})$ is assumed to be given by Eq.~(\ref{fullfact}).
Here we have taken the  conventions that $u(0,n)=0$ (the hop
rate is zero if there are no particles at the departure site),
$u(m,-1)=0$, and $m_0:= m_L,m_{L+1}:=m_1$.  Dividing both sides by 
$\prod_{i=1}^L f(m_i)$ and shifting
indices $(i-1,i)\to (i,i+1)$ in terms on the right-hand, we obtain
\bq
\sum_{i=1}^L \left[ u(m_i,m_{i+1}) -
  u(m_{i}+1,m_{i+1}-1)\frac{f(m_{i}+1)f(m_{i+1}-1)}{f(m_{i})f(m_{i+1})}
  \right] = 0. \label{balance2}
\eq
In Appendix A we show that the
most general solution to an equation  of the form
\bq \sum_{i=1}^L
F(m_i,m_{i+1}) =0
\eq
reads $F(a,b)=h(a)-h(b)$ for $L>2$. Therefore, from Eq.~(\ref{balance2}) we 
obtain that
\bq
u(m,n) - u(m+1,n-1)\frac{f(m+1)f(n-1)}{f(m)f(n)} = h(m)-h(n), \label{hmn}
\eq
with some (yet unknown) function $h(m)$. Let us consider now the case
  $n=0$. Because $u(m+1,-1)$ is identically zero, the second term in the above
  equation vanishes and we obtain
\bq
h(m)=u(m,0)+h(0). \label{hm}
\eq
Similarly, for $m=0$ and arbitrary $n$, the first term of Eq.~(\ref{hmn}) 
vanishes and we find
\bq
-u(1,n-1)\frac{f(1)f(n-1)}{f(0)f(n)} = h(0)-h(n) = -u(n,0). \label{hm0n}
\eq
The last relation (\ref{hm0n}) may be rearranged to give a recursion 
relation  for $f(n)$ as follows
\bq
f(n) = f(n-1) \frac{f(1)}{f(0)} \frac{u(1,n-1)}{u(n,0)}. \label{fn}
\eq
Inserting now (\ref{hm}) and (\ref{fn}) into Eq.~(\ref{hmn}) we arrive at 
the constraint on $u(m,n)$ which allows for the factorized steady state:
\bq
  u(m,n) = u(m+1,n-1)
  \frac{u(1,m)u(n,0)}{u(m+1,0)u(1,n-1)} + u(m,0) -
  u(n,0). \label{miscond}
\eq
It can be proved (see Appendix B) that this actually reduces to two 
conditions:
\ba
u(n,m) = u(m+1,n-1)
  \frac{u(1,m)u(n,0)}{u(m+1,0)u(1,n-1)}, \label{c1} \\
  u(n,m) - u(m,n) = u(n,0)-u(m,0). \label{c2}
\ea
These  conditions were first written down by Cocozza-Thivent \cite{CT85}
and can also be derived by using the condition of dynamical reversibility 
\cite{Pablo}.  Alternative proofs are presented in \cite{EH05} and
\cite{Godreche07}.

To obtain insight into the physical meaning of conditions 
(\ref{c1},\ref{c2}), let us examine the matrix $u_{mn}\equiv u(m,n)$. The 
first row  is  made up of zeros since $u_{0 n} = u(0,n)=0$.
Let us define two vectors $\vec{x}=\{x_1,x_2,\dots\}$ and 
$\vec{y}=\{y_1,y_2,\dots\}$  of infinite length which specify the
second row and the first column, i.e.,
\ba
u(m,0) = y_m, \label{ydef} \\
u(1,n) = x_n. \label{xdef}
\ea
The matrix $u_{mn}$ assumes now the following form:
\bq
u = \left[\begin{array}{ccccc}
0 & 0 & 0 & 0 & \dots \\
y_1 & x_1 & x_2 & x_3 & \dots \\
y_2 & . & . &. & \\
y_3 & . &.  &.  & \\
\vdots &  & & &
\end{array}\right].
\eq
Equations (\ref{c1}) and (\ref{c2}) can now be rewritten as two recursion 
relations which allow one to find $u(m,n)$ for $m>2,n>1$:
\ba
u(m+1,n-1) = u(n,m) \frac{y_{m+1}}{x_m} \frac{x_{n-1}}{y_n}, \label{c3} \\
u(n,m) = u(m,n) - y_m + y_n. \label{c4}
\ea
By iterating these equations one obtains unambiguous expressions for all 
$u(m,n)$,
for example, given $u(1,n)$ for $n>0$, (\ref{c4}) implies $u(n,1)$ and 
(\ref{c3}) gives $u(2,n-1)$.
Thus  the hop rate is fully determined by fixing the rate $y_m$ of hopping 
from a site with $m$ particles to an empty site, and the rate $x_n$ of 
hopping from a site with only one particle to a site with $n$ particles.
There is, however, an additional condition that $u(m,n)$ must be 
non-negative for all $m,n$. This imposes some constraints on $y_m,x_n$ which 
we were not able to express in a closed form and hence we do not have an
expression for the most general form of $y_m,x_n$ that implies non-negative 
$u(m,n)$. In what follows we will consider some special cases of $y_m,x_n$ 
for which one can prove non-negativeness of $u(m,n)$.

Let us now derive a formula for the weight function $f(n)$.  Iterating 
Eq.~(\ref{fn}) and using the definitions of $y_m,x_n$ 
(\ref{ydef},\ref{xdef}) we obtain
\bq
f(n) = f(0) \left(\frac{f(1)}{f(0)}\right)^n \prod_{i=1}^n 
\frac{x_{i-1}}{y_i}. \label{fngen}
\eq
It is important to note that any exponential factor $Aq^n$ in $f(n)$ does 
not change the steady-state properties since it appears in 
$P(m_1,\dots,m_L)$ as a constant prefactor $A^L q^{\sum_i m_i}=A^Lq^M$ due 
to the fixed number of sites, $L$, and  particles $M = \sum_i m_i$. Thus we 
may  neglect the factor  $f(0)(f(1)/f(0))^n$ in Eq.~(\ref{fngen}) and  write
\bq
f(n) =  \prod_{i=1}^n \frac{x_{i-1}}{y_i}. \label{fngen2}
\eq

\subsection{Lack of Pair-factorized states}

One may wonder whether conditions (\ref{c1},\ref{c2}) could be relaxed if, 
for example, instead of insisting on the factorization of the steady state 
over sites we assumed a much less-restrictive factorization over pairs of 
sites (Eq.~(\ref{pairfact})) as in Ref.~\cite{EHM06,WSJM09}. There, the hop 
rate depended on the occupation number $m_i$ of the departure site as well 
as two neighbouring sites $m_{i-1}, m_{i+1}$. Surprisingly, it turns out 
that for the misanthrope process, whose hop rate depends only on the 
departure and arrival site, pair-factorization is not possible except for a 
trivial case when the steady state fully factorizes as in 
Eq.~(\ref{fullfact}). We give a proof of this result in Appendix C.

\section{Physical examples of the hop rate}
In this section we present some particular  solutions to
(\ref{c1},\ref{c2}) or equivalently
(\ref{c3},\ref{c4}) which represent physical models, in the sense that the 
hop rates are non-negative.

\subsection{Factorized hop rate}
One can check that for a factorized hop rate of the form
\bq
u(m,n)=w(m)v(n),
\eq
equation (\ref{c1}) is automatically fulfilled. When $v(n)=\rm constant$, 
equation (\ref{c2}) is also fulfilled and we recover the ZRP with 
$u(m,n)=w(m)$. When $v(n)\neq \rm constant$, equation (\ref{c2}) leads to 
the relation between $w(n)$ and $v(n)$:
\bq
w(n) = C[v(n)-v(0)],
\eq
with some arbitrary, non-zero constant $C$. Thus
\bq
u(m,n) = C [v(m)-v(0)] v(n)  \label{ufact}
\eq
is the form of a factorised hop rate that yields a factorised steady state. 
For non-negativeness of $u(m,n)$, it suffices that one of the following 
conditions is fulfilled:
\begin{enumerate}
\item $C>0$, $v(m)>v(0)>0$ for $m>0$,
\item $C<0$, $v(0)> v(m) >0.$
\end{enumerate}
In both these cases $v(m)$ is always positive.
One could  additionally  consider cases where  $v(m)$
is always negative but as   $u(m,n)$ is invariant under $v \to- v$
these are trivially equivalent to the two cases above.

Our first simple example
of the factorised hop rate (\ref{ufact})
is $C=-1$ and $v(n)=N-n$ with some integer $N>0$. This leads to
\bq
u(m,n) = m(N-n).
\eq
If $N=1$ this reduces to the asymmetric simple exclusion process
where the occupancy of each site is limited to 1
and $u(1,1)=0$.
Similarly the case of general integer $N >0$
 corresponds to `partial exclusion'
\cite{SS} where each site of a lattice contains at most $N$ particles
and each particle  attempts hops forward to the next site
with rate one which succeed with probability $N-n$ where $n$ is the 
occupancy of the destination site.

The case $v(m) = m$ for $m>0$    has been studied and is  referred to as
the Inclusion Process. In a prescribed limit $v(0) \to 0$, the limit of zero 
hopping rate onto empty sites, a  form of condensation   occurs \cite{GRV}.

\subsection{Non-factorized  choices of $u(m,n)$}
The factorized form (\ref{ufact}) is not the only allowed form for $u(m,n)$
that solves (\ref{c3},\ref{c4}) and gives rise to a factorized steady state. 
To see this, let us consider an example in which $x_n=y_{n+1}$, with 
$y_n>0$. In this case, one can check that
\bq
u(m,n) = \sum_{i=0}^n y_{m+i} - \sum_{i=1}^n y_i
\eq
is the solution to equations (\ref{c3}), (\ref{c4}).
To ensure that the hop rate is positive, a sufficient condition is that 
$y_n$ grows with $n$.

Another interesting example is when $y_m=y=\rm const>0$ and $x_n$ is 
positive but otherwise arbitrary.
Then (\ref{c4}) implies that $u(n,m)=u(m,n)$ and (\ref{c3}) can be iterated 
to yield
for $m>0,n>1$
\bq
u(m,n) = \frac{\prod_{i=m}^{m+n-1} x_i}{\prod_{j=1}^{n-1} x_j}.
\eq
The corresponding  weight function (\ref{fngen2}) reads for $n>1$
\bq
f(n) = y^{-n}\prod_{i=1}^{n-1} x_i
  \to \prod_{i=1}^{n-1} x_i,\label{fn10}
\eq
where in the last expression we have suppressed a term exponential in $n$ 
which as, previously noted, is unimportant.
This expression assumes a form similar to the weight function in the 
zero-range process.
In fact, this model is closely related to a recently considered facilitated 
exclusion process \cite{GKR10}. In that work  an exclusion process is 
considered wherein, for a particle to hop, the  neighbouring site behind has 
to be occupied. Using the standard mapping of section 1 (see 
Fig.~\ref{mapping_to_tasep}) the facilitated exclusion process corresponds 
to the limit $u(m,0) = y = 1$ and $u(m,n) = x \to 0$ of the misanthrope 
process.

\section{Condensation in the misanthrope process\label{seccond}}
An important feature of models such as ZRP \cite{EH05} or balls-in-boxes 
model \cite{BBJ}, in which the   steady-state probability takes the 
factorized form (\ref{fullfact}), is that for some choices of the weight 
function $f(n)$ the system exhibits a phase transition in the thermodynamic 
limit $M,L\to\infty$. If the density  of particles $\rho=M/L$ is above some 
critical value $\rho_c$, the surplus of particles $M-\rho_c L$ accumulates 
at a single site and is called the condensate. The sufficient condition for 
the condensation above some finite $\rho_c$ is the appropriate asymptotic 
behaviour of $f(n)$ \cite{EH05}.
As any exponential dependence in $f(n)$ is irrelevant (see previous 
section), we just need to consider the asymptotic  behaviour of $f(n)$
modulo any exponential factor. There are two generic cases (see e.g. 
\cite{EMZ06}):
\begin{description}
\item[I] $f(m)\sim m^{-\gamma}$  with $\gamma>2$. The critical density is 
finite but its numerical value depends on the particular form of $f(m)$ and 
not only on its asymptotic behaviour. The fraction $\rho/\rho_c-1$ of all 
particles goes to the condensate. We will refer to this as standard 
condensation.

\item[II] $f(m)$ increases with $m$ more quickly than exponentially, e.g., 
as $\sim m!$. This leads to so called strong (or complete) condensation - 
the critical density $\rho_c=0$ and a fraction of particles
tending to one in the thermodynamic limit
is located at one  site.
\end{description}
There are also some other specific  examples, such as the Backgammon  model 
\cite{backgammon,backgammon2},
which corresponds to taking $f(m)=e^{\beta \delta_{m_i,0}}$
and exhibits condensation in the limit $\beta\to\infty$,  and  the inclusion 
process in the limit $v(0)\to 0$ mentioned earlier \cite{GRV}.

To see why the condensation happens in the two  generic cases highlighted 
above, we shall follow a standard approach \cite{EH05}. Treating the 
steady-state probability as the statistical weight of a given configuration, 
and defining the grand-canonical partition function
\bq
G(z) = \sum_{\{m_i\}} 
\prod_{i=1}^L f(m_i) z^{m_i} = F(z)^L,
\eq
where
\bq
F(z) = \sum^{\infty}_{m=0} f(m) z^m, \label{Fz}
\eq
we see that the phase transition, signalled by a singularity  of $G(z)$ at 
some $z_c$, is possible only if the series $F(z)$
has a finite radius of convergence $z_c$. Moreover, the density calculated 
as a function of fugacity $z$ from the grand-canonical partition function,
\bq
\rho = z \frac{F'(z)}{F(z)},
\label{rho}
\eq
must be finite as $z\to z_c^{-}$. This is only possible if $f(m)$ decays as 
a power law in which case we may have a finite $\rho_c$ (case I) or $f(m)$ 
grows very fast with $m$ in which case $z_c=0$ and $\rho_c=0$ (case II). Any 
exponential factor in $f(m)$ only shifts the position of $z_c$.

\subsection{Factorized hop rate}
We will now discuss which choices of the hop rate $u(m,n)$ of the 
misanthrope process  lead to the  generic cases {\bf I, II} described above.
We begin by considering factorized  hop rates $u(m,n)$,  as in 
Eq.~(\ref{ufact}). Using
\bq
x_n=v(n)(v(1)-v(0)), \qquad y_m=v(0)(v(m)-v(0)), \label{factxy}
\eq
equation~(\ref{fngen2}) yields
\bq
\frac{f(m)}{f(m-1)} \propto  \frac{v(m-1)}{v(m)-v(0)}, \label{ratiofm}
\eq
where we have removed a constant factor that generates an exponential factor 
in $f(m)$.
Let us first consider  strong condensation (case {\bf II}), that is 
$f(m)/f(m-1)\to \infty$ for large $m$. From Eq.~(\ref{ratiofm}) we see that 
for this  to happen, $v(m)$ has to tend to $v(0)$ in the limit of 
$m\to\infty$. For example, for
\bq
v(0)=1, \qquad v(m)=1+\frac{1}{m+1}, \label{ex1}
\eq
we obtain $f(m)/f(m-1) = (m+1)^2/m$ and strong condensation occurs.

In standard condensation (case {\bf  I}) we require a power law decaying as
\bq
f(m)\sim m^{-\gamma},\label{fmg}
\eq
so that
\bq
\frac{f(m)}{f(m-1)} \sim 1- \frac{\gamma}{m}.\label{fm1g}
\eq
Expression (\ref{ratiofm}) fulfils this condition for two distinct  large 
$m$, asymptotic behaviours of $v(m)$:
\bq
v(m) \sim m^\gamma, \label{cc1}
\eq
and
\bq
v(m) \cong \beta \left(1-\frac{\alpha}{m}\right). \label{cc2}
\eq
Clearly, the different hop rates (\ref{cc1},\ref{cc2}) constructed from  the 
different choices for $v(m)$
will lead to very different dynamical properties system which
we shall study  in detail in   Section \ref{sec_dynamics}.

For the first case (\ref{cc1}), condensation is possible for any $\gamma>2$. 
The exact form of $v(m)$ is not important. For the second case (\ref{cc2}), 
we obtain from Eq.~(\ref{ratiofm}) that
\bq
\frac{f(m)}{f(m-1)} \simeq  A \left( 1- \frac{\alpha v(0)}{m(v(0)-\beta)} 
\right),
\eq
where $A$ is a constant. Therefore the asymptotic behaviour of the 
single-site weight is
$f(m) \simeq A^m m^{-\gamma}$ where the exponent $\gamma$  reads
\bq
\gamma = \alpha \frac{v(0)}{v(0)-\beta}.
\eq
Condensation is possible if $\gamma$ is larger than 2. This leads to the 
following conditions  on $\beta$, $v(0)$ and $\alpha$ for condensation to 
occur:
\ba
\mbox{for}\; \beta <v(0) &:& \alpha>\frac{2(v(0)-\beta)}{v(0)}, \\
\mbox{for}\; \beta >v(0) &:& \alpha<- \frac{2(\beta- v(0))}{v(0)}. \label{bga}
\ea
Interestingly, the {\em existence} of condensation depends not only on the 
asymptotic behaviour of $v(m)$ but also on $v(0)$. This should be contrasted 
with the ZRP  for which it is  only the asymptotic decay of the hop rate,
given by $\alpha$ in 
$u(m)\cong\beta(1+\alpha/m)$, that determines whether condensation is 
possible or not and the exact form of $u(m)$ affects only the value of the 
critical density, above which condensation occurs. Here, the influence of 
the particular form of  $v(m)$ is much more significant.

\subsection{Illustrative example of a factorized hop rate}
As an illustrative example consider the hop rate
\bq
v(0)<1, \qquad v(m)=1+\frac{1}{m+1}, \label{ex2}
\eq
which differs from Eq.~(\ref{ex1}) only in the value of $v(0)$.
In the previous case, $v(0)=1$  we have strong condensation, whereas 
$v(0)<1$ in  Eq.~(\ref{ex2}) corresponds to $\alpha=-1,\beta=1$ from 
Eq.~(\ref{cc2}).
Thus, by (\ref{bga}),  for $v(0)\leq 2/3$  there is no condensation, but for $2/3< v(0) <1$ there 
is standard condensation  and for $v(0)=1$ there is strong condensation, 
even though $v(m)$ is the same in all cases for $m>0$.

It is possible to compute exactly the critical density in this case by using 
(\ref{fngen2}), which yields
\bq
f(n) = \frac{(n+1)! (n+1)!}{n! (c)_n},
\eq
after removing  exponential factors in $n$.
Here
\bq
c=\frac{3-2v(0)}{1-v(0)},
\eq
and we use the Pochhammer symbol  $(c)_n = c (c+1)\ldots (c+n-1)$.
Then the generating function (\ref{Fz}) is a hypergeometric function
\bq
F(z) = \sum^{\infty}_{n=0} \frac{(n+1)!\, (n+1)!}{n!\, (c)_n} z^n = 
{}_2F_1(2,2,c\,;z),
\eq
which converges for $z \leq 1$.
Using standard identities for the hypergeometric function (which hold when 
the series converge) \cite{Andrews},
\ba
{}_2F_1(a,b,c\,;1) = 
\frac{\Gamma(c)\Gamma(c-a-b)}{\Gamma(c-a)\Gamma(c-b)}, \\
\frac{\partial }{\partial z}\, {}_2F_1(a,b,c\,;z) = \frac{ab}{c}
{\,}_2F_1(a+1,b+1,c+1\,;z),
\ea
one may determine the critical density as the maximum allowed density 
(\ref{rho}),
given by $z \to 1$:
\bq
\rho_c = \left.\frac{z F'(z)}{F(z)}\right|_{z=1} = \frac{4}{c} \frac{ 
{\,}_2F_1(3,3,c+1\,;1)}{ {\,}_2F_1(2,2,c\,;1)} = \frac{4}{c-5} = 
\frac{4(1-v(0))}{3v(0) -2}.
\eq
We see indeed that $\rho_c \to 0$ as $v(0)\to 1$
and $\rho_c \to \infty$ as $v(0)\to 2/3$.

\subsection{General case with arbitrary $x_m,y_n$}
We now turn to the case of general $u(m,n)$ satisfying (\ref{c1},\ref{c2}).
From Eq.~(\ref{fngen}) we obtain that
\bq
\frac{f(m)}{f(m-1)} = \frac{x_{m-1}}{y_m}.
\eq
In order to have  strong condensation (case {\bf II}) we require 
$x_{m-1}/y_m\to\infty$ for
large  $m$. For standard condensation  with $f(m)\sim m^{-\gamma}$ 
(\ref{fmg})
(case {\bf I})  we have again two distinct possible  asymptotic behaviours 
for $x_m,y_n$:
\bq
y_n \cong 1 + \frac{A}{n}, \qquad x_m \cong 1+ \frac{B}{m}, \label{ccc1}
\eq
which gives $\gamma=A-B$, and $A-B>2$ as the necessary condition for 
condensation, and
\bq
y_n \sim (n+C)^\alpha, \qquad x_m \sim (m+D)^\alpha, \label{ccc2}
\eq
which in turn leads to $\gamma=\alpha(1+C-D)$. The condition for 
condensation reads now $\alpha(1+C-D)>2$. As before, the two asymptotic 
forms of hop rates (\ref{ccc1}) and (\ref{ccc2}) lead to different dynamics.

\section{Dynamics of the condensate}
\label{sec_dynamics}
We return  now to the different microscopic dynamics that yield the same
steady state and exhibit condensation above the same $\rho_c$.
To gain insight into the dynamics, we simulated this system using a modified
continuous time  Monte Carlo  in which the time interval to the next event 
is stochastically generated. We prepare the system in a state in which 
particles are randomly distributed among the sites. Then, in each time step 
we choose a random site $i$ and move a particle from this site to its right 
neighbour. The site is selected at random with probability $u(m_i,m_{i+1}) 
/\sum_j u(m_j,m_{j+1})$. To do this effectively, we sort all possible 
``moves'' with respect to their probabilities of occurrence. Lastly, we 
update the time $t\to t+\Delta t$ by adding an exponentially distributed 
random variable $\Delta t$ with mean $1/\sum_j u(m_j,m_{j+1})$. To minimize 
computational time, after each move we recalculate only the rates 
$u(m_{i-1},m_i)$, $u(m_i,m_{i+1})$, and $u(m_{i+1},m_{i+2})$ which have been 
affected by the move. We also use tabulated values of $u(m,n)$ to avoid 
time-consuming algebraic operations such as calculating $m^\gamma$ for 
non-integer $\gamma$.

\subsection{ZRP-like rates given by Equation (\ref{cc2})}
\begin{figure}
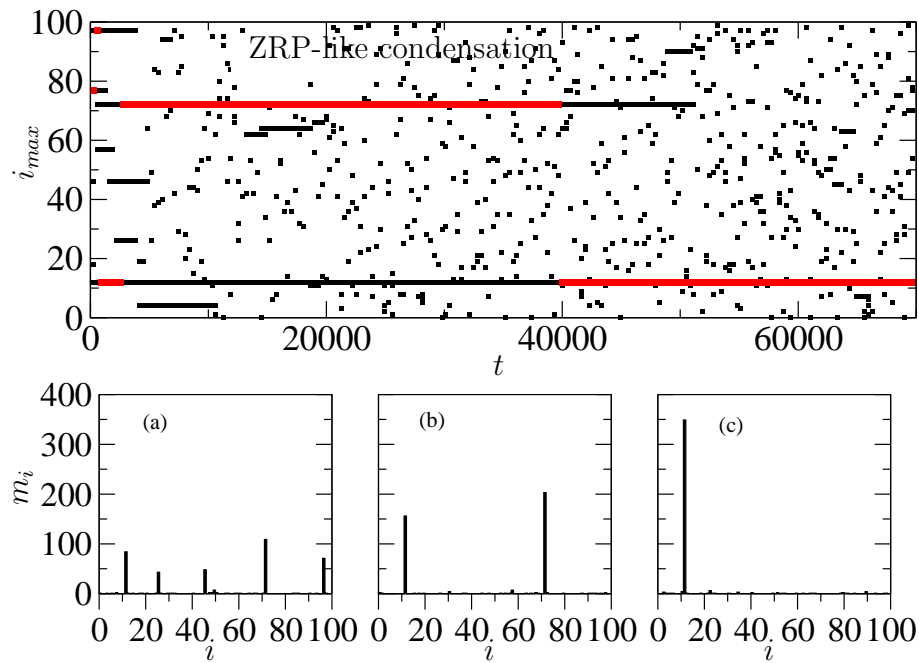

\psfrag{y1}{$i_{max}$}\psfrag{x1}{$t$}
\psfrag{xx}{$i$}\psfrag{yy}{$m_i$}
\psfrag{zrp}{ZRP-like condensation}
\includegraphics*[width=12cm]{dynamics_normal.eps}\\
\includegraphics*[width=12cm]{dynamics_normal_snapshots.eps}
\caption{Top: positions of five most occupied sites as a function of time 
(squares) in a system with $L=100$ sites, $M=400$ particles, and the 
factorized hop rate with $v(m)=\beta \left(1-\frac{\alpha}{m}\right)$ 
defined in Eq.~(\ref{cc2}) for $v(0)=1.3,\alpha=0.7,\beta=1$ (this 
corresponds to $\gamma\approx 3.03$). Red squares show the position of the 
site with the maximal number of particles (the condensate). Bottom: three 
snapshots of the system for a) $t=2800$, b) $t=30000$, c) $t=80000$. After a 
fast nucleation stage during which many small condensates are formed, the 
condensates slowly merge into a single one.}
\label{dyn_norm}
\end{figure}

\begin{figure}
\includegraphics*[width=7cm]{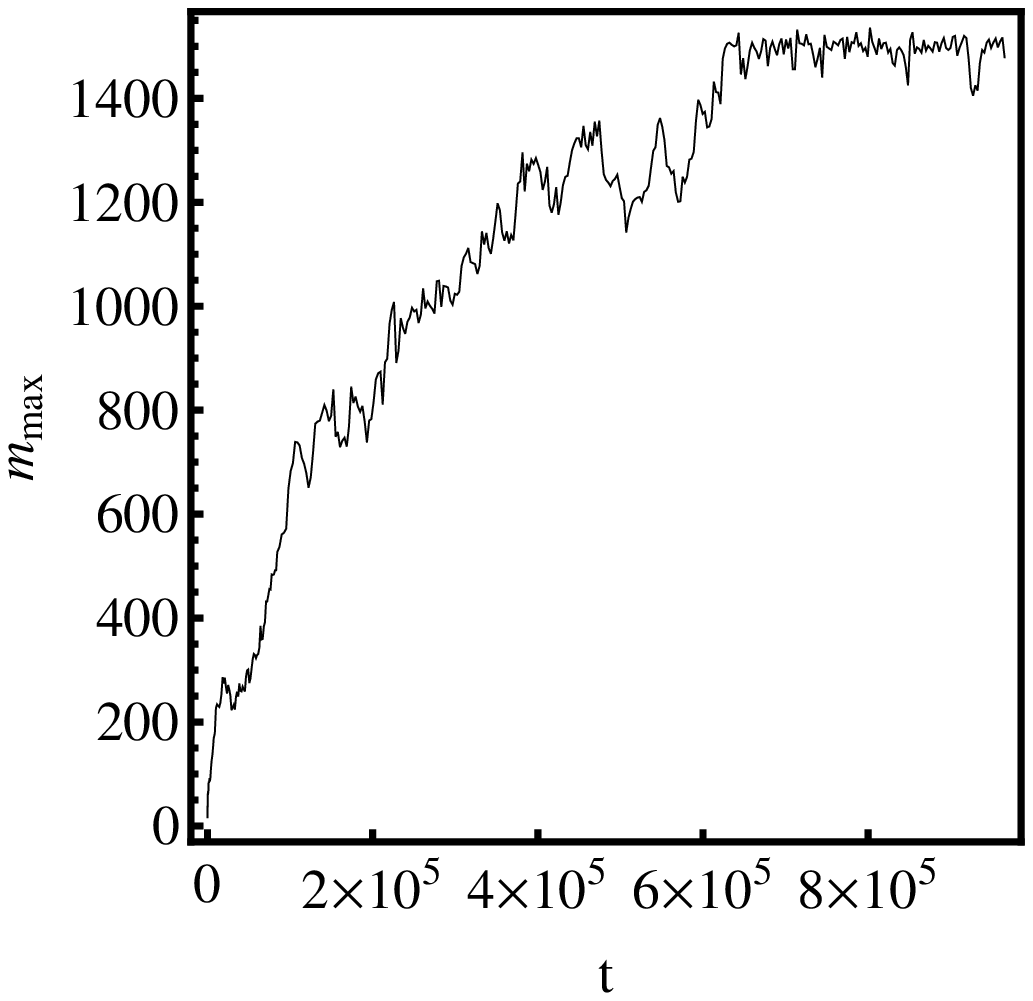}
\includegraphics*[width=7cm]{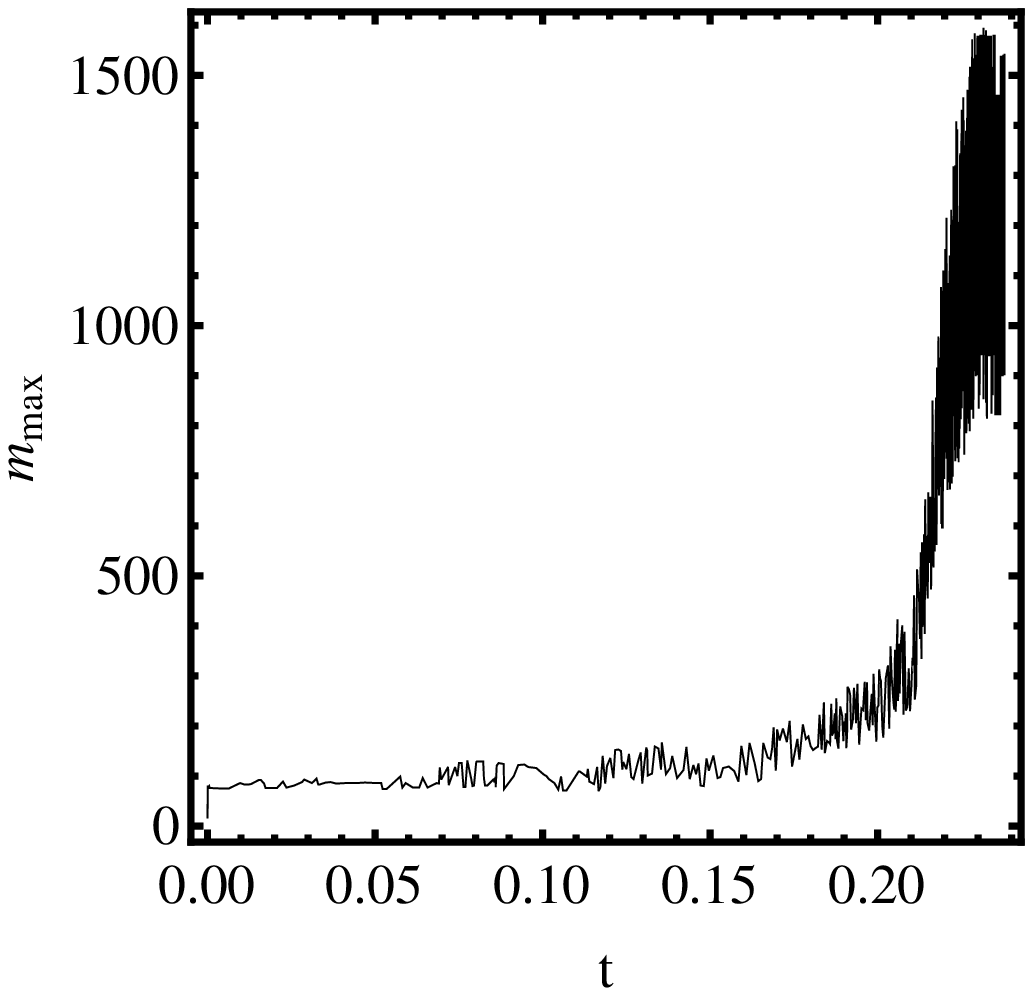}
\caption{\label{fig:mmax}Simulation runs show the occupation $m_{\rm max}$ 
of the site with maximal number of particles as a function of time $t$. 
Left: ZRP-like condensation ($v(0)=1.3,\alpha=0.7,\beta=1$ in 
Eq.~(\ref{cc2})), right: explosive condensation with $v(m)=(m+0.1)^3$. 
$L=200$, $M=1600$ (density $\rho=8$) in both cases.}
\end{figure}

In Figure \ref{dyn_norm} we show a graph for a typical simulation run for 
ZRP-like hop rates (\ref{cc2}), that is
\bq
v(m) = \beta\left(1 -\frac{\alpha}{m}\right)\quad\mbox{for}\quad m>0,
\eq
and $v(0)>\beta$. Then taking $C=-1$ in (\ref{ufact}), yields the asymptotic 
behaviour
\bq
u(m,n) \cong \beta (v(0)-\beta) - \frac{\alpha\beta (v(0)-\beta)}{n} + 
\frac{\alpha\beta^2}{m}, \label{umn_powlaw}
\eq
for $m,n$ sufficiently large. In this figure, the positions of the five 
most-occupied sites are plotted against time, together with snapshots of the 
system at three different times. We see that at first, small condensates are 
formed (multiple horizontal lines in the upper diagram). These condensates 
coalesce quickly until only two are left. Finally, the two condensates merge 
into a single one; this last process is the slowest. The size of the 
condensate initially grows quickly and then slows down, see 
Fig.~\ref{fig:mmax}, left. This is very similar to the condensation in the 
zero-range process with $u(m)=1+\gamma/m$, which also leads to the power-law 
$f(m)\sim m^{-\gamma}$. The dynamics of such ZRP has been extensively 
studied in the past \cite{Godreche03,GL05}. It has been argued that the 
typical time to reach the steady state in an asymmetric 1d system scales as 
$L^2$ with the number of sites $L$ if the density $\rho=M/L$ is kept 
constant and $M,L$ are sufficiently large.

\begin{figure}
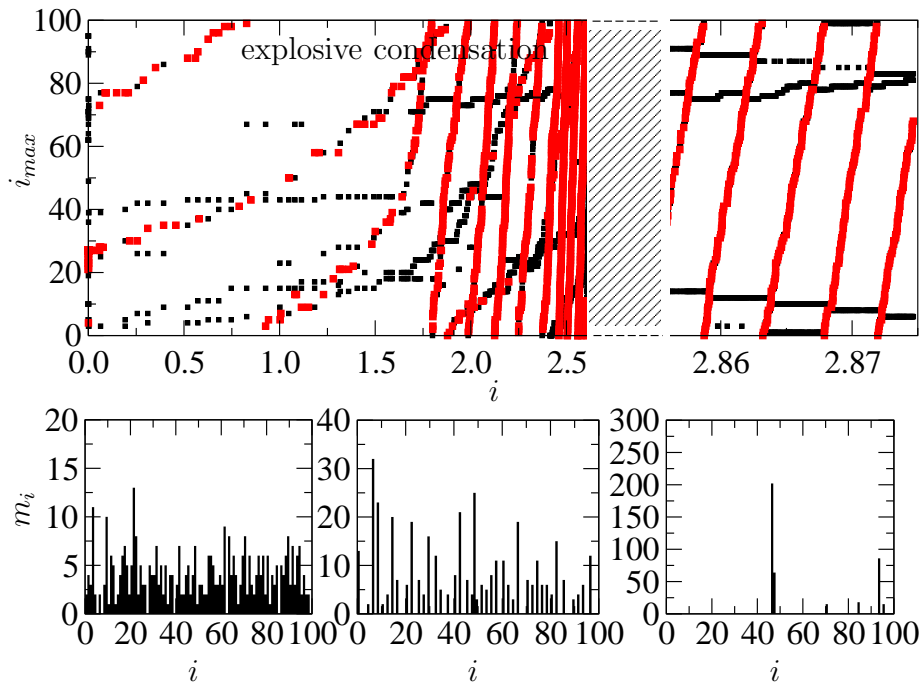

\psfrag{yy}{$i_{max}$}\psfrag{x1}{$t$}
\psfrag{xx}{$i$}
\psfrag{our_model}{explosive condensation}
\includegraphics*[width=12cm]{dynamics_explosive.eps}\\
\psfrag{yy}{$m_i$}
\includegraphics*[width=12cm]{dynamics_expl_snapshots.eps}
\caption{Top: positions of five most occupied sites (as in 
Fig.~\ref{dyn_norm}) as a function of time (squares) in a system with 
$L=100$ sites, $M=400$ particles, and the factorized hop rate 
$v(m)=(m+0.1)^3$ as in Eq.~(\ref{cc1}) for $\gamma=3$. Red squares show the 
position of the condensate. Bottom: three snapshots of the system for a) 
$t=0$, b) $t\approx 1$, c) $t\approx 2.86$. Initial nucleation is the 
slowest process, followed by a rapid growth of the condensate which ``sucks 
up'' particles as it moves at increasing velocity to the right. The velocity 
of the condensate can be read off from the slope of the red line in the top 
figure. The terminal velocity of the condensate in the steady state (not 
shown here) is $\approx (0.1M)^3\approx 64000$.}
\label{dyn_expl}
\end{figure}

One can argue that the same scaling should hold in our system, that is the 
$n$-dependence of $u(m,n)$ in Eq.~(\ref{umn_powlaw}) is not important. The 
argument goes as follows. In the last, slowest stage of condensation when 
only two condensates are left, occupation numbers on sites surrounding these 
condensates fluctuate very rapidly. Thus, from each condensate's point of 
view, the hop rates $u(m_{i-1},m_i), u(m_i,m_{i+1})$ to  and from the 
condensate at site $i$ are effectively averaged over $m_{i-1},m_{i+1}$. 
According to formula (\ref{umn_powlaw}), the average inflow rate is 
$\left<u(m_{i-1},m_i)\right> \cong {\rm const} - \frac{\alpha\beta 
(v(0)-\beta)}{m_i}$ while the outflow rate reads 
$\left<u(m_{i},m_{i+1})\right> \cong {\rm const} + 
\frac{\alpha\beta^2}{m_i}$. The net flow between both condensates having 
$m_1$ and $m_2$ particles, respectively, is $\sim (1/m_1 - 1/m_2)$. Since 
both condensates have $O(L)$ particles, the flow is at most $O(L^{-1})$. 
Thus, statistical fluctuations which are of order $u(m_i,m_{i+1}) \sim 
u(m_{i-1},m_i) \sim O(1)$ dominate and $m_1,m_2$ perform effectively 
unbiased random walks, at least as long as they are of order $O(L)$. The 
typical time scale is defined by the time it takes to reach either $m_1=0$ 
or $m_2=0$, which scales diffusively as $T\sim L^2$. The finite distance 
$O(L)$ between the condensates does not matter, because due to the finite 
average hop rate in the bulk, the particles traverse it in $\sim L$ steps 
which is smaller than the time scale related to fluctuations.

\begin{figure}
\psfrag{xx}{$L$}\psfrag{yy}{$T$}
\includegraphics*[width=12cm]{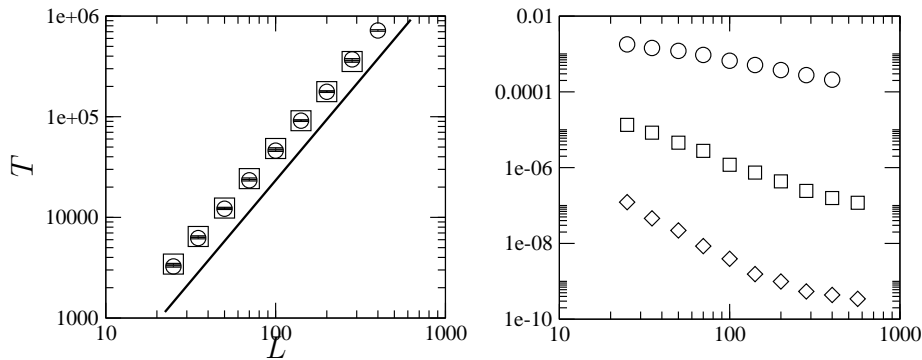}
\caption{Left: time to reach the steady state as a function of the system 
size $L$, for $\rho=4,\beta=1,\alpha=0.7$ and for $v(0)=1.3$ (circles) and 
$v(0)=1.21$ (squares), which correspond to $\gamma\approx 3,\rho_c\approx 
0.5$ and $\gamma\approx 4,\rho_c\approx 0.3$, respectively. Thick line 
corresponds to a theoretical prediction $T\sim L^2$. Right, the same plot 
for $v(m)$ from Eq.~(\ref{eq:umneps}) (explosive condensation) with 
$\epsilon=1$ and particle density $\rho=8$. Curves from top to bottom: 
$\gamma=3,4,5$.
}
\label{times}
\end{figure}

To confirm this prediction, we carried out simulations for $\rho\gg \rho_c$ 
and various sizes $L$, and measured the time $T$ it took before the maximal 
occupation number reached the mean steady-state value $M-L\rho_c$. One 
clearly sees from Fig. \ref{times}, left, that $T$ scales approximately as 
$\sim L^2$, independently of $\gamma$ and other parameters.

\subsection{Explosive-condensation rates given by Equation (\ref{cc1})}
We now turn to the case $v(m)\sim m^\gamma$ from Eq.~(\ref{cc1}), for which 
$u(m,n) \sim (mn)^\gamma$ for large $m,n$. To be more specific, we choose 
$v(m)=(\epsilon+m)^\gamma$ so that
\bq
u(m,n) = ((\epsilon+m)^\gamma-\epsilon^\gamma)(n+\epsilon)^\gamma , 
\label{eq:umneps}
\eq
with (typically) $\epsilon\ll 1$.
In Fig.~\ref{dyn_expl} we show results of computer simulations.
If we compare it with Fig.~\ref{dyn_norm}, we see some striking differences. 
First, the site carrying the maximal number of particles moves 
unidirectionally, in the same direction as hopping particles. The motion of 
the condensate is similar to the ``slinky''-like motion of a
non-Markovian model~\cite{HMS09,HMS12} or a generalization of ZRP to 
non-factorizing probabilities \cite{HMS12v2}. Second, the evolution speeds 
up in time; the condensate moves faster and faster as it gains particles, 
see also Fig.~\ref{fig:mmax}, right. This is why in the previous work 
\cite{prl} we termed it  ``explosive condensation''. Third, the speed at 
which the condensate travels through the system stabilizes after the system 
reaches the steady state. Finally, other (smaller) peaks move in the 
opposite direction to the main condensate each time they interact with it.

The dynamics thus differs significantly from the ZRP-like case. Figure 
\ref{times}, right, shows another striking feature of this new condensation. 
If we fix the density of particles $\rho$ and increase the system size $L$, 
the time $T_{\rm ss}$ to reach steady state {\it decreases} as a function of 
$L$. This is in contrast to the ZRP-like behaviour from Fig.~\ref{times}, 
left, and to the ordinary ZRP, where the time to steady state increases with 
$L$. In Ref.~\cite{prl} we focused on deriving the relation between $T_{\rm 
ss}$ and $L$, assuming that the time scale $T_{\rm ss}$ was dominated by the 
slow process of initial coalescence, in contrast to the ZRP where this 
process is the fastest one. Our reasoning was based on several assumptions 
on how clusters of particles interact, which we inferred from numerical 
simulations. In what follows we shall provide more evidence in support of 
these arguments.

First, we show that a cluster of particles of size $m$ moves through the 
system at speed $\approx (\epsilon m)^\gamma$. As it moves, the cluster 
collides with other clusters and exchanges particles with them. On average, 
particles flow from smaller to larger clusters. Next, we show that the rate 
of accumulation increases with increasing speed of a cluster and, as a 
result, it takes a finite time for the cluster to reach macroscopically 
large occupation (condensate). We also show that the distribution of this 
time is quite broad and arbitrarily short times are possible (although not 
very likely). Finally, we show that, since there are initially $O(L)$ 
clusters and each of them can potentially become the condensate in a finite 
time, the time to reach steady state (condensation) decreases with $L$ as a 
result of a simple extreme value statistics argument.

We shall now discuss our arguments in detail, assuming that $\epsilon \ll 
1$. Let us first calculate the speed of an isolated cluster of $m$ particles 
travelling in an otherwise empty system. We assume that the cluster is 
initially at site $i$, so that $m_i=m$, and $m_{i+1}=0$. The only allowed 
transition for the cluster is to lose one particle to site $i+1$, which 
happens with rate $u(m,0)$. The average time for this to happen is 
$t_m=1/u(m,0)$. In the second step, the condensate may lose another particle 
with rate $u(m-1,1)$, which will take time $t_{m-1}=1/u(m-1,1)$. 
Alternatively, the particle at site $i+1$ may jump to site $i+2$, but this 
is much less likely because $u(m_{i+1},0)\propto \epsilon^\gamma\ll 1$ and 
this is much smaller than $u(m_i,m_{i+1})$. The process in which site $i$ 
loses particles in favour of site $i+1$ will thus continue with rate 
$u(m_i,m-m_i)$ (with other processes having negligible rates) until $m_i=0$ 
and $m_{i+1}=m$. The average times for each step will be 
$t_m,t_{m-1},\dots,t_1$, where $t_n = 1/u(n,m-n)$. Therefore, the total time 
$\tau$ of the process will be the sum of times $t_m,t_{m-1},\dots,t_1$:
\bq
\left<\tau\right> = \sum_{n=1}^m t_n = \sum_{n=1}^m \frac{1}{u(n,m-n)} = 
\sum_{n=1}^m \frac{1}{(m-n+\epsilon)^\gamma [(\epsilon+n)^\gamma - 
\epsilon^\gamma]}.
\eq
For $\gamma>2$ which is the case of condensation and for $\epsilon\ll 1$, we 
can approximate the above sum by the last term only, and we obtain that 
$\left<\tau\right>\approx (\epsilon m)^{-\gamma}$. This shows that a cluster 
with $m$ particles moves $1/\left<\tau\right> \approx (\epsilon m)^\gamma$ 
sites per unit time. In particular, the speed of the condensate of size $M 
\approx (\rho-\rho_c) L$ is $(\epsilon(\rho-\rho_c))^\gamma L^\gamma$ and 
increases as a power of the system size $L$. Although in our analysis we 
have neglected a small probability of losing particles by the cluster as it 
moves through the system, this is justified for both small clusters and the 
macroscopic cluster (the condensate). For small clusters of size $1\ll m\ll 
M$ (initial stage of condensation), the distance between the clusters is of 
order $O(\rho L/m)$ and the probability of losing a particle between 
collisions $O(\epsilon^\gamma L/m)$ is negligible. For the condensate 
($m\approx M$), it is possible to lose particles but as it moves through the 
system, the condensate will also gain particles from the background and its 
size will fluctuate only a little. The above formula is in excellent 
agreement with simulations; for example, for parameters from 
Fig.~\ref{dyn_expl}, the measured speed is $63850\pm 100$ sites per unit 
time, whereas the formula $(\epsilon M)^\gamma$ gives $64000$.

\begin{figure}
\includegraphics[width=12cm]{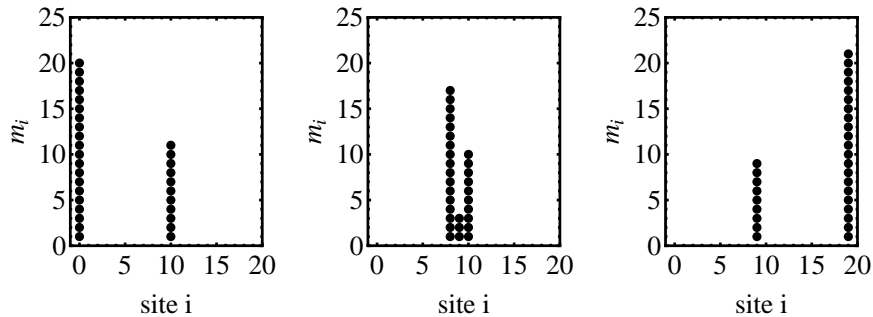}
\caption{\label{fig:coll_setup}Example of an ``experimental'' setup for 
studying collisions of condensates. A cluster of size $m_0=20$ collides with 
a smaller cluster of size $n_0=10$ placed at $i=L/2=10$. Pictures from left 
to right: initial state, collision, and final state at which the mass 
$\Delta m$ is measured at site $i=L-1=19$.}
\end{figure}

\begin{figure}
\includegraphics[width=4cm]{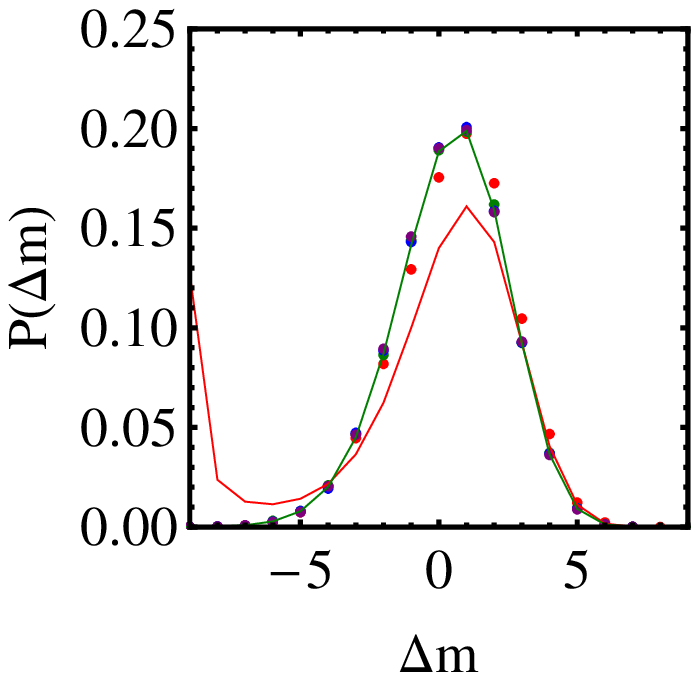}
\includegraphics[width=4cm]{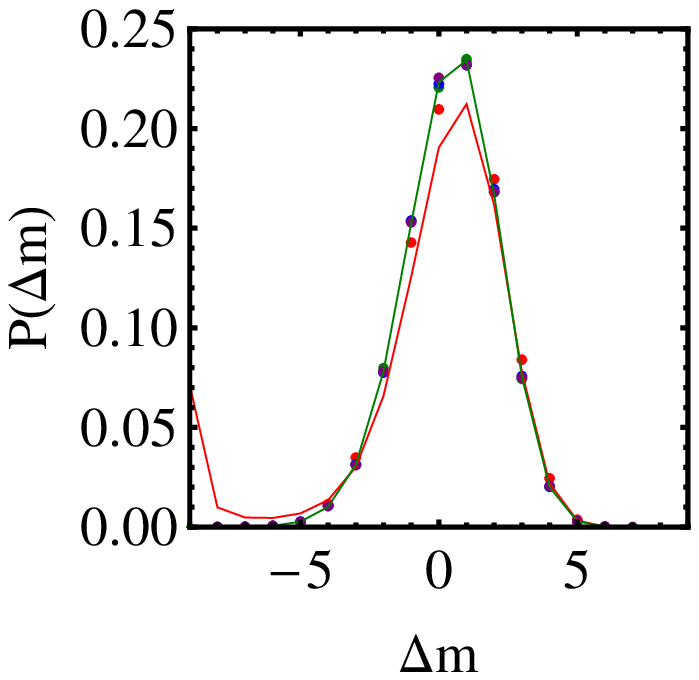}
\includegraphics[width=4cm]{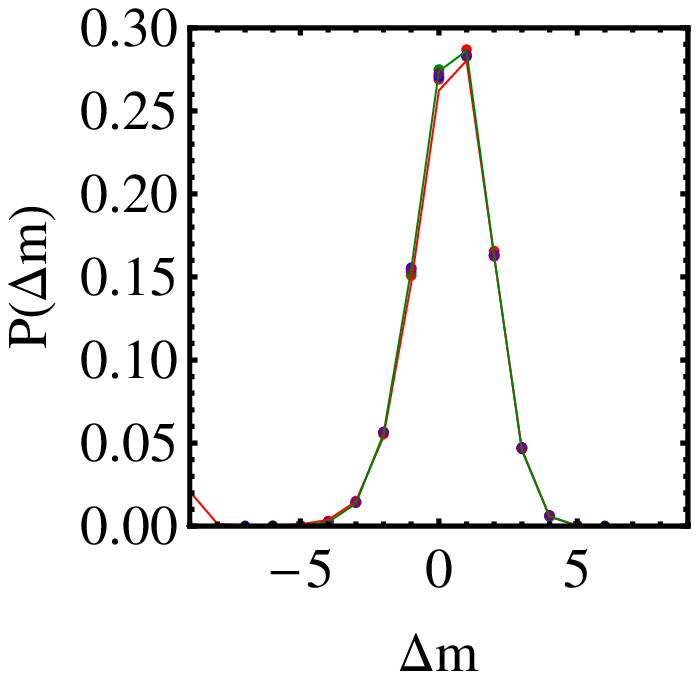}
\caption{\label{fig:Pdm}Plots of $P(\Delta m)$ for $\gamma=3$ (left), 
$\gamma=4$ (middle), and $\gamma=6$ (right). Points correspond to MC 
simulations for $n_0=10$ and $m_0=20,50,100,200$ (red, green, blue, and 
purple, respectively), whereas continuous lines are calculated using 
recursion~(\ref{eq:rec}) for $n_0=10$ and $m_0=20,50$ (red and green, 
respectively).}
\end{figure}

We shall now analyse what happens in a collision between two clusters of 
size $m_0$ and $n_0$. The process can be conveniently studied in numerical 
simulations by performing ``scattering experiments'' in which a larger 
condensate $m_0$ collides with a slower, smaller condensate $n_0$, see 
Fig.~\ref{fig:coll_setup}. We prepare the system in a state such that $m_0$ 
and $n_0$ are initially located at sites $i=0$ and $i=L/2$, respectively, 
and we simulate the system using the same kinetic Monte-Carlo algorithm as 
before until one of the condensates reaches site $i=L-1$. We then measure 
its mass $m_0'$ and calculate the mass $\Delta m$ transferred in any 
collision\footnote{In principle, no collision may occur but the probability 
of this is vanishingly small for large enough $L$} from the smaller to the 
larger condensate: $\Delta m = m_0' - m_0$. In Fig.~\ref{fig:Pdm} we show 
that although the distribution of $\Delta m$ looks almost symmetrical, it is 
biased toward positive values. The average $\left<\Delta m\right> \approx 
0.4$ and depends only weakly on the size of the clusters involved in the 
collision and on the value of $\gamma$. The average mass transfer does not 
also depend on the value of $\epsilon$, although $\epsilon>0$ is necessary 
to have a non-zero hop rate for a single particle. Thus, in our further 
analysis below, we shall assume $\epsilon\to 0$ whenever possible as this 
greatly simplifies calculations.

\begin{figure}
\includegraphics[width=8cm]{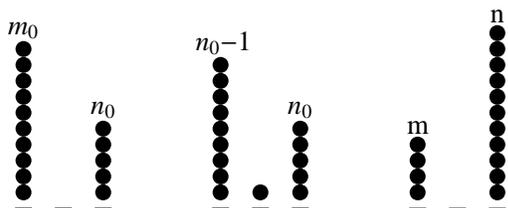}
\caption{\label{fig:collision}Collision between two clusters of sizes $m_0$ 
and $n_0$ (left picture) can be approximately reduced to the dynamics of 
only three sites. The collision is initiated when a particle hops from the 
left cluster to the middle site (middle picture). The collision is finished 
when the middle site reaches zero occupation (right picture).}
\end{figure}

\subsubsection{Deterministic description of scattering}
\begin{figure}
\includegraphics[width=12cm]{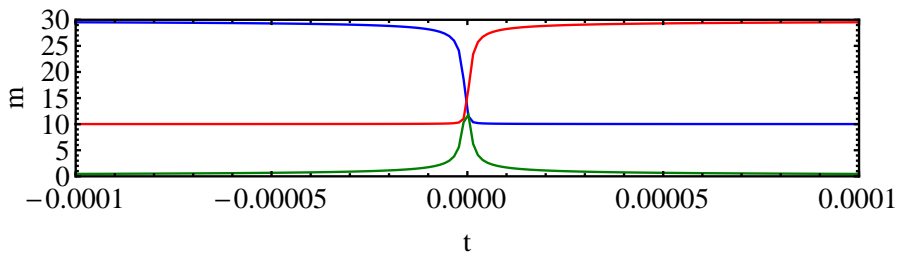}
\caption{\label{fig:coll_determ}Solution of 
Eqs.~(\ref{eq:mav})-(\ref{eq:nav}) for $m_0=30-\epsilon_2, n_0=10$ and 
$M=m_0+n_0=40$. Curves: $m(t)$ (blue), $n(t)$ (red), and the middle-site 
occupation $M-m(t)-n(t)$ (green). Time axis has been shifted so that $t=0$ 
corresponds to the peak of the middle-site occupation $M-m-n$. }
\end{figure}

To understand the numerical results, we assume (as before) that the system 
is completely empty, save for the two clusters. As the larger cluster $m_0$ 
approaches the smaller cluster $n_0$, the collision is initiated when $m_0$ 
and $n_0$ become separated by just a single site with one particle that has 
hopped out from $m_0$, see Fig.~\ref{fig:collision}. The initial 
configuration is therefore: $m=m_0-1$ particles at the left site, $1$ 
particle at the middle site, and $n=n_0$ particles at the right site. In the 
collision process, the left site loses particles to the middle site, which 
in turn loses particles in favour of the right site, with the transition 
rates given by $u(m,M-m-n)$ and $u(M-m-n,n)$, respectively, where 
$M=m_0+n_0$ is the total number of particles in the system (this is a 
conserved quantity). As a first step, it is useful to describe the dynamics 
of this process using a deterministic approach in which stochasticity is 
neglected. Our expectation would be that this description should be valid 
for large clusters, for which random fluctuations of $m,n$ do not matter. 
The equations for the average $m,n$ read
\ba
\dot{m} = -u(m,M-m-n) = -[m(M-m-n)]^\gamma, \label{eq:mav} \\
\dot{n} = u(M-m-n,n) = [n(M-m-n)]^\gamma, \label{eq:nav}
\ea
with the initial condition $m(0)=m_0-1,n(0)=n_0$. We note that
\bq
\frac{\dot{m}}{m^\gamma} + \frac{\dot{n}}{n^\gamma} = 0,
\eq
which means that the quantity
\bq
m^{1-\gamma} + n^{1-\gamma} = \rm const
\eq
is a conserved quantity determined by the initial condition. We also note 
that upon time reversal $t\to {\rm const}-t$, the variables $m$ and $n$ 
exchange their roles:
\ba
\dot{n} = -[n(M-m-n)]^\gamma, \\
\dot{m} = [m(M-m-n)]^\gamma,
\ea
and thus if we change $m\to n$ and $n\to m$, we will recover 
Eqs.~(\ref{eq:mav})-(\ref{eq:nav}). This means that the solution 
$(m(t),n(t))$ for $t\to\infty$ must be equal to the initial condition at 
$t=0$, and hence $m(t\to\infty) = n_0$, $n(t\to\infty) = m_0-\epsilon_2$ 
(where $\epsilon_2$ is small). In Fig.~\ref{fig:coll_determ} we plot 
numerical solutions of Eqs.~(\ref{eq:mav})-(\ref{eq:nav}) which confirm our 
prediction. The deterministic, continuous approximations predicts zero mass 
transfer because the size of the larger cluster is the same as before the 
collision. This is at variance with the results of stochastic simulations, 
even for very large $m,n$, hence we see that stochasticity is important.

\subsubsection{Stochastic description of scattering}
\begin{figure}
\includegraphics[width=8cm]{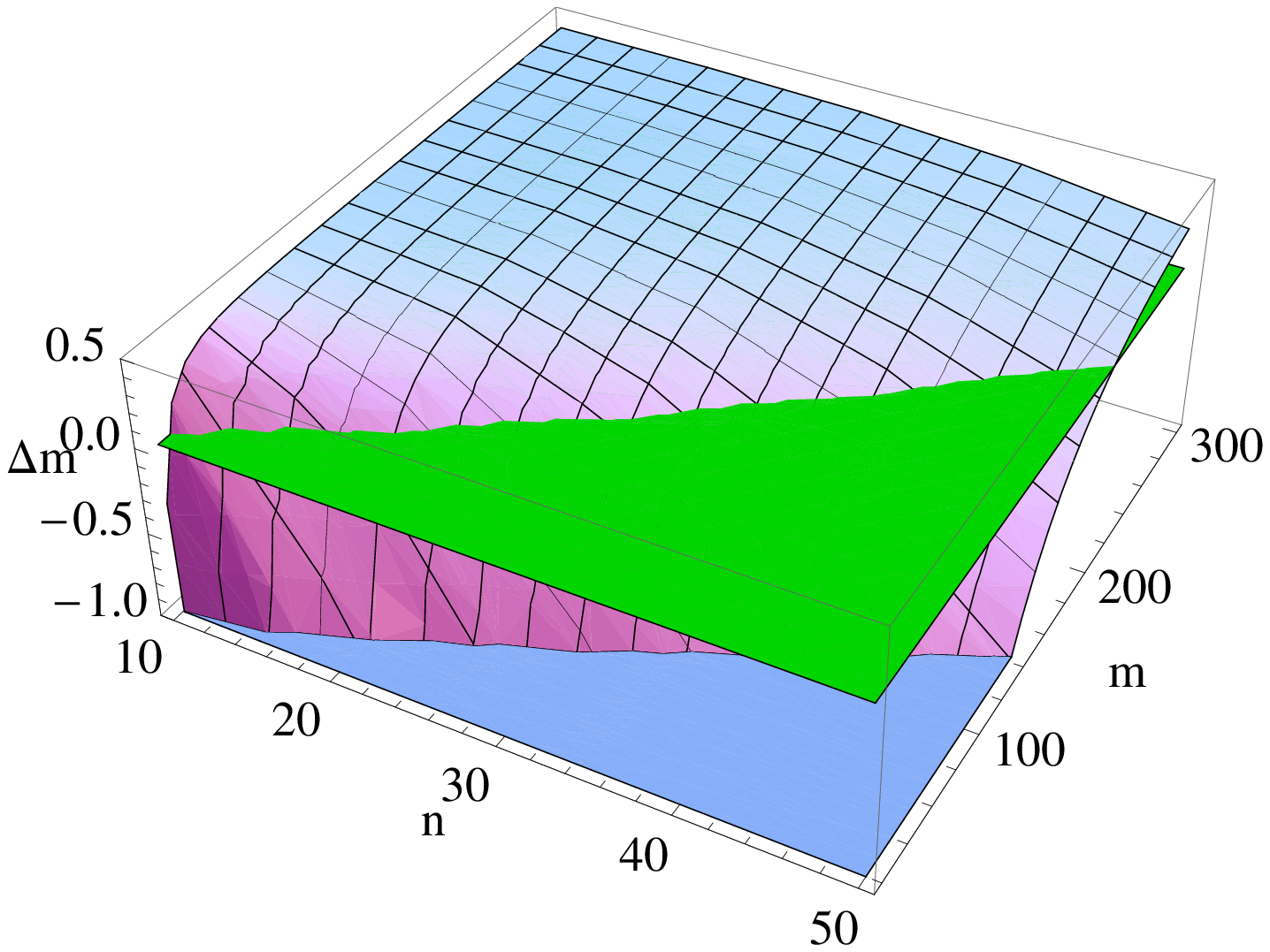}
\includegraphics[width=8cm]{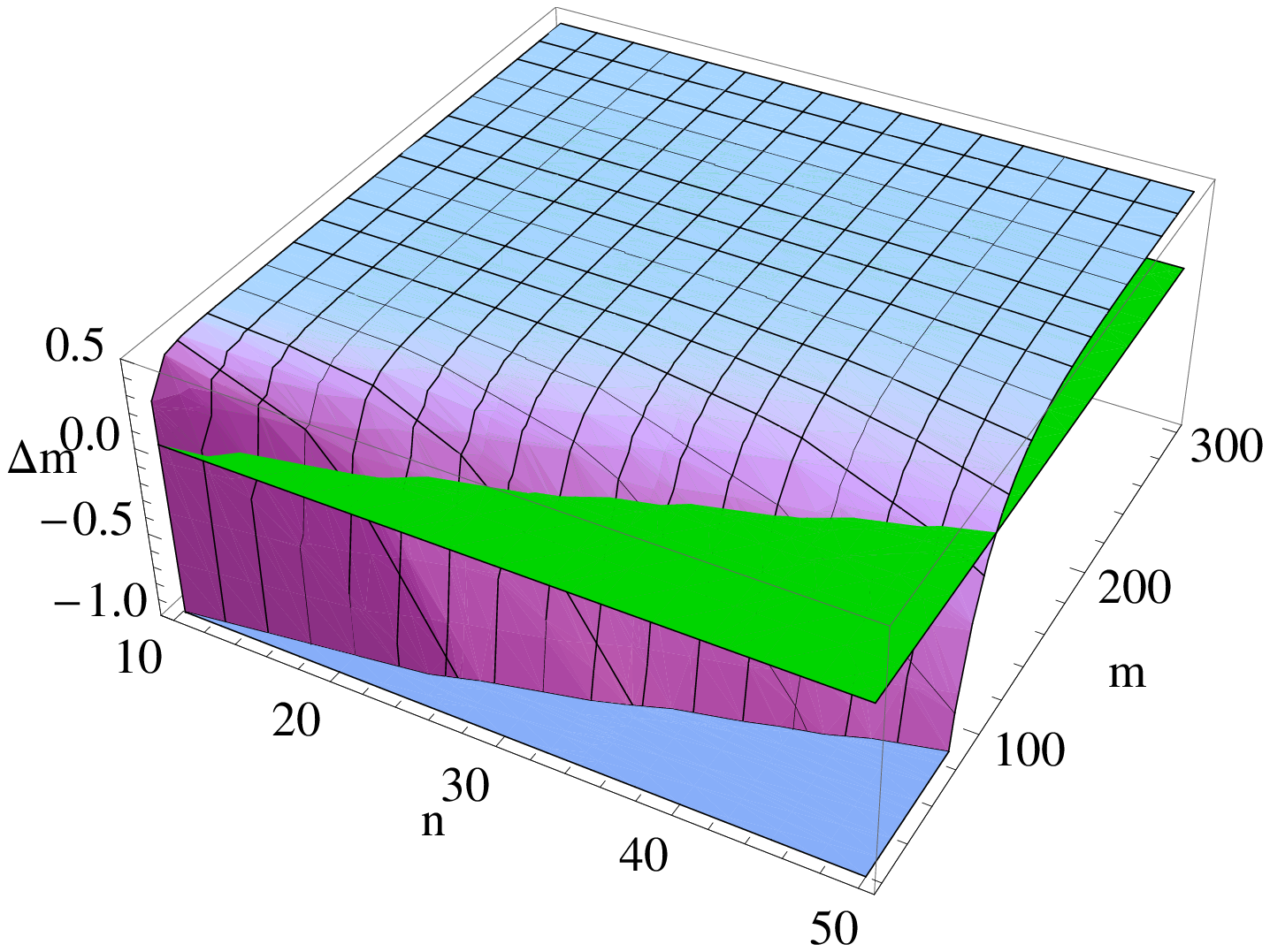}
\caption{\label{fig:dm_stoch}Plots of the average mass $\Delta m$ 
transferred in a collision of clusters of size $m_0,n_0$, calculated using 
recursion (\ref{eq:rec}). Left: $\gamma=4$, right: $\gamma=6$. The 
half-transparent plane corresponds to $\Delta m=0$. The range of $\Delta m$ 
has been restricted to $-1,\dots,0.5$ for clarity, although for some 
$m_0,n_0$, $\Delta m$ takes values less than $-1$.}
\end{figure}

\begin{figure}
\includegraphics[width=8cm]{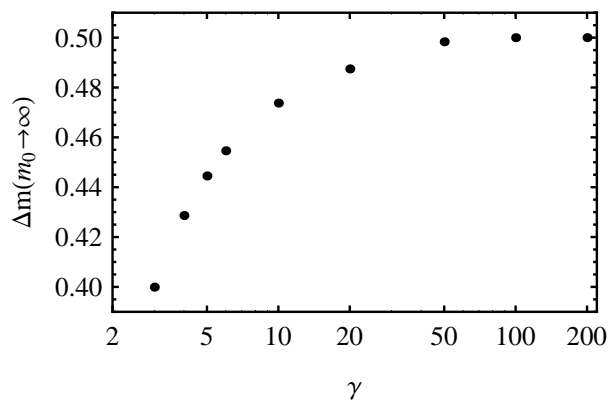}
\caption{\label{fig:dm_limit}Plot of $\Delta m$ for $m_0\to\infty$ and 
$n_0=10$ as a function of $\gamma$. Transferred mass tends to $0.5$ for 
large $\gamma$.}
\end{figure}

It is convenient to express the evolution of the fully stochastic problem 
not in  real time $t$, but as a function the number of particles $k$ that 
have hopped since the beginning of the process. Assume that $k=1$ 
corresponds to the state $(m_0-1,1,n_0)$. From the deterministic problem and 
Fig.~\ref{fig:coll_determ} we know that the occupation $M-m-n$ of the middle 
site first rises and then decreases again. The collision ends when the 
middle site reaches zero occupation, so that the final state is 
$(m,0,m_0+n_0-m)$ with $m$ a random variable.
The mass transferred  from the smaller to the larger condensate  in the 
collision is given by $\Delta m = (m_0+n_0-m)-m_0 = n_0-m$.
Let us denote by $P(m,n;k)$ the probability that the state of the three 
sites is $(m,M-m-n,n)$ after $k$ hops. To calculate the mass transferred in 
the process, we need to calculate
\ba
\left<\Delta m\right> = \sum_{m=0}^{m_0-1} (n_0 - m) P(m, m_0 + n_0 - m; 
2(m_0 - m) - 1) \\
= \sum_{\Delta m=n_0-m_0+1}^{n_0}  \Delta m P(\Delta m), \label{eq:dm}
\ea
where $P(\Delta m)= P(n_0-\Delta m, m_0 + \Delta m;  2(m_0 - n_0 + \Delta 
m) - 1)$ is the distribution of the mass transferred. The probability 
$P(m,n;k)$ can be found recursively:
\ba
P(m,n;k) = \frac{u(m+1,M-m-n-1)P(m+1,n;k-1)}{u(m+1,M-m-n-1)+u(M-m-n-1,n)} 
\nonumber \\
+ \frac{u(M-m-n+1,n-1)P(m,n-1;k-1)}{u(m,M-m-n+1)+u(M-m-n+1,n-1)} \nonumber 
\\
= \frac{(m+1)^\gamma P(m+1,n;k-1)}{(m+1)^\gamma+n^\gamma}  + 
\frac{(n-1)^\gamma P(m,n-1;k-1)}{m^\gamma+(n-1)^\gamma}, \label{eq:rec}
\ea
with the initial condition $P(m,n;0) = \delta_{m,m_0-1} \delta_{n,n_0}$. In 
Fig.~\ref{fig:dm_stoch} we plot $\Delta m$ calculated numerically using the 
above formula and Eq.~(\ref{eq:dm}), for different sizes $m_0,n_0$ of 
colliding condensates, and $\gamma=4$. We can see that for $m_0\gg n_0$, the 
mass $\Delta m$ transferred in the collision tends to a constant value of 
about $0.4-0.5$, depending on $\gamma$. In Fig.~\ref{fig:dm_limit} we show 
that $\Delta m$ stabilises at $0.5$ for $m_0\to\infty$ and 
$\gamma\to\infty$. We also see in Fig.~\ref{fig:dm_stoch} that for 
$m_0\approx n_0$, the average mass calculated from the recursion relation is 
actually negative, as if particles flew from the larger to the smaller 
condensate. This does not agree with Monte-Carlo simulations, see 
Fig.~\ref{fig:Pdm}, which indicate that $\Delta m\approx \rm const$, even 
for a small difference between $m_0,n_0$. Since our numerical calculation is 
exact, this discrepancy implies that in the simulation, clusters collide 
more than once, until the right cluster becomes sufficiently large compared 
to the left cluster, and manages to escape.

\subsubsection{Coalescence process of clusters}
Our simulations and numerical calculations thus indicate that we can 
describe the relaxation towards steady state as a process of coalescence of 
small clusters into the condensate in a series of collisions. In each 
collision, only a few particles are exchanged on average, and the net flow 
of particles is from smaller to larger clusters. Let us analyse what happens 
if the system is initially prepared in a state with $O(L)$ clusters 
separated by runs of empty sites. Numerically, this can be realised by 
placing $N=L/W$ clusters at sites $i=1,W,2W,\dots$, where $W$ is some fixed 
number larger than 1 and independent of $L$. This initial state is not the 
same as the state with randomly distributed particles that we typically use 
in simulations, but the existence of well-defined clusters at all times 
makes the effective description from the previous subsections (in terms of 
collisions between isolated clusters) applicable to the whole process, from 
the initial to the final state. Moreover, on a large scale the system still 
looks homogeneous, as if the particles were distributed uniformly.

\begin{figure}
\centering
\includegraphics*[width=6cm]{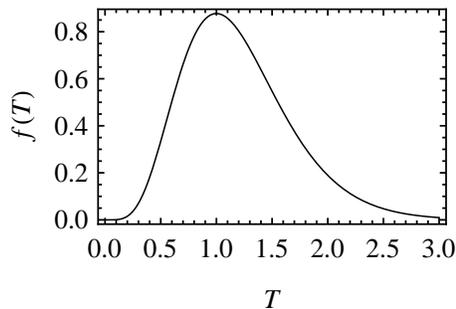}
\caption{\label{fig:exft}A hypothetical example of the function $f(T)$ 
which gives the probability that a randomly chosen cluster will become the 
condensate. The time to steady state is determined by the behaviour of 
$f(T)$ for $T\to 0^+$.}
\end{figure}

Let us first look at the evolution of one cluster. It collides with other 
clusters, gains or loses particles, and finally it either dissolves into the 
background or it reaches the steady-state occupation $M-L\rho_c$ and becomes 
the condensate. Let $T$ be the time it takes to reach this steady-state size 
for this particular condensate ($T=\infty$ if the cluster dissolves into the 
background). $T$ differs among different clusters and different realisations 
of the system, and it can be thought of as a random variable with some 
distribution $f(T)$. In Figure \ref{fig:exft} we show a hypothetical form of 
such a distribution. Since all clusters are initially identical, $f(T)$ is 
the probability density function of a randomly chosen cluster having evolved 
into
a condensate after time $T$. We have, however, $N$ such clusters and each of 
them should have the same chance to become the final condensate, although 
only one of them (the fastest evolving one) will make it. Therefore, the 
time $T_{\rm ss}$ to reach steady state will be the minimal time out of 
times $T_1,\dots,T_N$ for individual clusters:
\bq
T_{\rm ss} = \min \left\{ T_1,\dots, T_N \right\}, \label{tss}
\eq
and we can estimate it as $T_{\rm ss}$ such that\footnote{In Ref.~\cite{prl} 
we give a more complicated argument also based on extreme value statistics, 
which leads however to the same final result.}
\bq
\int_0^{T_{\rm ss}} f(T) dT = 1/N. \label{eq:tssf}
\eq
This argument is based on statistical independence of $T_1,T_2,\dots,T_N$. 
The rationale behind it is that each cluster can be viewed as moving in a 
random background formed by other clusters. Since we are interested in 
large-$N$ (and therefore large-$L$) behaviour, we need only the small-$T$ 
behaviour of the single-cluster distribution $f(T)$.

\begin{figure*}
\includegraphics[width=15cm]{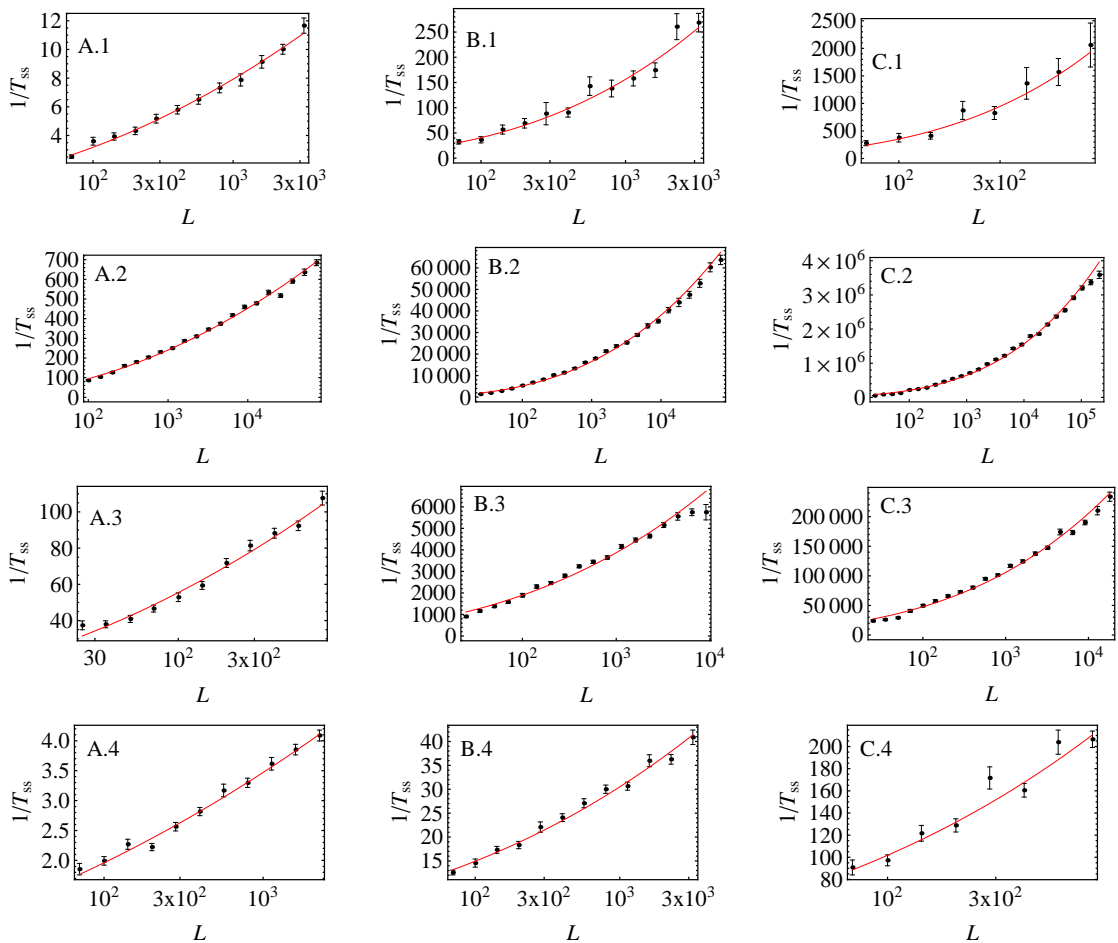}
\caption{\label{fig:allplots}Plots of $T_{\rm ss}^{-1}$ as a function of $L$ 
(points). Solid line is Eq.~(\ref{Tfit}) fitted to the data points. Columns 
A-C: $\gamma=3,4,5$, from left to right. Rows 1-4 correspond to 1) 
Poissonian initial distribution of particles, $v(m)=(m+0.3)^\gamma$, 2) 
Poissonian distribution, $v(m)=(m+1)^\gamma$, 3) Poissonian distribution 
with only every 5th site occupied, $v(m)=(m+1)^\gamma$, 4) the same but 
$v(m)=(m+0.3)^\gamma$.}
\end{figure*}

Let us first, quite generically, write equations for the evolution of mass 
$m_n$ and time $t_n$ at which a particular cluster moves by one site to the 
right and (possibly) exchanges a chunk of mass $\Delta m_n$ with other 
clusters:
\ba
m_n &=& m_{n-1} + \Delta m_n, \\
t_n &=& t_{n-t} + \Delta t_n.
\ea
Here $\Delta m_n$ is a random number that can be either zero (the cluster 
moves to the next site without changing its mass), positive, or negative. We 
expect that the distribution of $\Delta m_n$ should be similar to the 
distribution $P(\Delta m)$ from Fig.~\ref{fig:Pdm}. $\Delta t_n$ is the time 
between two jumps and it is exponentially distributed:
\bq
p_n(\Delta t_n) = \lambda_n e^{-\lambda_n \Delta t_n}, \label{pndt}
\eq
where $\lambda_n$ is the rate constant, and we expect that $\lambda_n 
\propto m_n^\gamma$ for large $m_n$. The reason is that for the hop rate 
(\ref{eq:umneps}) and small $\epsilon$ the limiting process is the 
Poissonian process in which the first particle hops from the condensate to 
the next, empty site. We now want to calculate the distribution $f(T)$ of 
the sum $T=t_1+t_2+\dots$ since this is the time to reach size $m_n=\infty$ 
(condensation). In Ref.~\cite{prl} we show that $T$ is a random variable 
with the following distribution:
\bq
f(T)  \propto C T^{(1-3\gamma)/(2(\gamma-1))} \exp\left[-B 
(AT)^{-\frac{1}{\gamma-1}}\right], \label{f(T)}
\eq
where $B,C$ are some real, positive constants. If we now insert 
Eq.~(\ref{f(T)}) into Eq.~(\ref{eq:tssf}) and recall that $N\sim L$, we 
obtain that the relaxation time asymptotically decreases as
\bq
T_{\rm ss} = c_1 (\ln L + c_2 \ln\ln L + \dots)^{1-\gamma}, \label{Tfit}
\eq
where dots ``$\dots$'' denote terms increasing slower than $\ln\ln L$. For 
very large $L$, equation (\ref{Tfit}) predicts that the average time to 
reach the steady state should decrease as $(\ln L)^{1-\gamma}$. The 
coefficients $c_1,c_2$ could be in principle calculated, but since our 
theory uses many approximations whose errors would accumulate in $c_1,c_2$, 
it makes more sense to treat them as free parameters when comparing 
Eq.~(\ref{Tfit}) with numerical simulations. This is what is done in 
Fig.~\ref{fig:allplots}, where we plot $1/T_{\rm ss}$ obtained in 
simulations, together with Eq.~(\ref{Tfit}) with $c_1,c_2$ fitted to the 
data points. The agreement is very good and it extends beyond the 
assumptions of the above theory. For example, the formula (\ref{Tfit}) works 
not only for the initial condition with $N$ isolated clusters of equal size, 
but also when the clusters have different numbers of particles, or when the 
initial condition is a random distribution of particles. Moreover, the 
formula works also when $\epsilon\approx 1$ is not necessarily a  small 
number.

Finally, we note that the effective description does not invoke a factorised 
steady state or a factorised hop rate.
Although our numerical studies presented in Section 5  have used a 
factorised hopping rate, we checked that similar results are obtained for a 
choice of dynamics that does not correspond to a factorised hop rate 
(nor satisfy the conditions for a factorised steady state), but has the same
asymptotic behaviour $u(m,n)\sim (mn)^\gamma$.

\section{Explosive condensation as a spatially-extended example of 
instantaneous gelation.\label{sec:instgel}}
As we have seen an effective description of  explosive condensation
which occurs for $v(m)\sim m^\gamma$
is given in terms of  coalescence of clusters through collisions.
In turn this description  bears some resemblance to  a coagulation process. 
These processes describe aggregation of clusters of particles that can be 
anything from atoms to stars, and have applications to such diverse problems 
as gravitational clustering \cite{silk-white}, formation of droplets in 
clouds due to collisions \cite{falkovich}, and differential sedimentation 
\cite{horvai}. What all these processes have in common is that the rate at 
which clusters merge and form larger clusters increases with their size. A 
particularly simple example is the growth of raindrops that fall through the 
mist under their own gravity. Since the terminal velocity of a raindrop 
increases with its size, and the rate of mass accretion will be proportional 
to the volume ``swept'' by the drop, the drop will aggregate mass at an 
increasing rate as it grows \cite{raindrops}.
In essence this is the common feature  with  explosive condensation 
described in the previous section.

The process of irreversible cluster-cluster aggregation is usually described 
by Smoluchowski equation \cite{Connaughton}:
\bq
\frac{d N_i}{dt} = \frac{1}{2} \sum_{j+k=i} K(j,k)N_j N_k - \sum_j K(i,j) 
N_i N_j.
\eq
Here $N_i$ is the number of clusters of size $i$. The function $K(i,j)$, 
which gives the rate at which clusters of sizes $i$ and $j$ merge, is called
 the coagulation kernel. In theoretical studies of aggregation processes, it 
is often assumed that the kernel is a homogeneous function,
\bq
K(ai,aj) = a^\lambda K(i,j), \label{kaij}
\eq
where $\lambda$ is the degree of homogeneity. Many kernels that arise from 
practical applications are homogeneous, so this restriction is very mild. 
One  form for  a kernel satisfying  (\ref{kaij}) is
\bq
K(i,j)=i^\nu j^\mu + i^\mu j^\nu, \label{kernel}
\eq
with $\nu>\mu$ and the degree of homogeneity $\lambda=\mu+\nu$. Although in 
the limit of infinitely large systems the final state---a single, infinite 
cluster---is the same regardless of $\nu,\mu$, the dynamics of this 
``gelation'' process (where the infinite cluster plays the role of gel) 
depends on $\lambda$ and $\nu$. For $\lambda<1$, the size of the largest 
cluster (gel) grows as a power of time, and gelation takes infinitely long. 
For $\lambda=1$, gelation still takes infinite time but the size of the gel 
grows exponentially. For $\lambda>1$, however, gelation takes finite time 
for $\nu<1$, or is believed to happen in zero time $\nu>1$ 
\cite{Connaughton, dongen, leyvraz}. This latter process is called 
``instantaneous gelation''.

Instantaneous gelation bears a close resemblance to  our explosive 
condensation,  with time $T_{\rm ss}$ to steady state decreasing as $(\ln 
L)^{1-\gamma}$ and hence becoming zero for $L\to\infty$. However, there are 
also important differences between our model and instantaneous gelation. 
First, clusters can split apart in our model, hence aggregation is 
reversible (albeit this occurs rarely for large clusters). The process of 
merging of clusters is also different to that of the Smoluchowski equation; 
our clusters do not completely merge upon contact but only exchange 
particles, with a net flow of particles from smaller to larger cluster. This 
is not a serious difference; we could imagine rescaling the kernel so that 
the aggregation rate would correspond to the same effective net flow. 
However, since in our model the rate of collisions is proportional to the 
relative velocity of clusters $i,j$, the effective exchange kernel would 
have to assume the following form:
\bq
K(i,j) \sim i^\gamma - j^\gamma,
\label{kernel2}
\eq
for $i>j$, and $K(i,j)=0$ otherwise. This is quite different to 
Eq.~(\ref{kernel}) which is symmetric in $i,j$. However, in the limit  $i\gg 
j$, which is relevant when one of the clusters begins to dominate, kernel 
(\ref{kernel2}) reduces to Eq.~(\ref{kernel}) with $\nu=\gamma$. 
Interestingly, in this regime our model becomes similar to exchange driven 
growth \cite{bn-krapivsky} in which particles are transferred between 
clusters of size $i$ and $j$ at a rate $K(i,j)$ given by Eq.~(\ref{kernel}). 
The model \cite{bn-krapivsky} leads to instantaneous gelation for $\nu>2$, 
which corresponds to $\gamma>2$ in our model. However, the fact that our 
effective kernel (\ref{kernel2}) is not symmetric in $i,j$ makes a 
difference in the scaling of the time to steady state. Specifically, our 
model predicts $T_{\rm ss} \sim (\ln L)^{-(\gamma-1)}$, whereas exchange 
driven growth \cite{bn-krapivsky} gives $T_{\rm ss} \sim (\ln 
L)^{-(\gamma-2)}$.
Finally we should bear in mind that the explosive condensation occurs on a 
spatially extended one-dimensional lattice whereas  coagulations processes 
are  essentially mean field models defined without any dimensionality

\section{Conclusion \label{secconc}}
In this work we have considered the factorised steady states in
the misanthrope process and the condensation phenomena that may
occur. We have seen that for the standard condensation scenario (case
{\bf I} of section 4) there are distinct type of dynamics that will yield
the same required asymptotic behaviour of the single-site weight
$f(m)$. This is in contrast to the ZRP. We have also
seen how the existence of condensation may depend not only on the
asymptotic behaviour of the hop rates but also on the behaviour of the hop 
rate for small $m$; specifically, condensation may or may not be possible, 
depending on the value $v(0)$ of
the function $v(m)$ introduced in section 3.1.

This  result may be of significance for example to the phase separation 
criterion proposed for driven diffusive systems \cite{KLMST02,ELMM04}.
There it was assumed that one could model domains in the driven system as 
sites in a zero range process  and relate the current out of a domain as the 
hopping rate of the  zero range process (depending only on the domain size).
Phase separation in the driven system then corresponds to condensation in 
the Zero range process.
If the current between two neighbouring domains has a more  complicated
relationship  between the two domain sizes, a description in terms of a 
misanthrope process might become more appropriate.

We have also analysed the dynamics of condensation and showed that if $v(m)$ 
increases as a power of $m$ (cf. equation (\ref{cc1})), the time to steady 
state decreases with increasing system size $L$. This means that in the 
limit $L\to \infty$, the steady-state condensate is produced 
instantaneously. This is at variance with ZRP-like dynamics (\ref{cc2}) and 
resembles instantaneous gelation known from the theory of coagulation 
processes discussed in Sec.~\ref{sec:instgel}. However, in contrast to these 
models with mean-field dynamics, our misanthrope process provides a 
non-trivial example of  instantaneous gelation in a spatially-extended 
system. Since our analysis is quite generic and is not based on the 
factorization of the steady state, we expect that the same dynamics will 
hold in any non-equilibrium system with hop rates increasing with occupation 
numbers.

Although our analysis of the dynamics is well confirmed by numerical 
simulations, it remains to be seen if a seemingly simple expression 
(\ref{Tfit}) for the time to condensation $T_{\rm ss}$ can be derived 
without various approximations that we have assumed. In particular, it would 
be interesting to see whether the same scaling $T_{\rm ss}\sim (\ln 
L)^{1-\gamma}$ remains true for an arbitrary initial condition.

\section*{Acknowledgments}
We would like to thank D. Mukamel, S. Grosskinsky, and C. Connaughton for 
helpful discussions. B.W. was supported by a Leverhulme Trust Early Career 
Fellowship. This work was funded in part by the EPSRC under grant number
EP/J007404/1,

\section*{Appendix A}
We shall show that the most general form of a function $F(a,b)$ which obeys 
the equation
\bq
\sum_{i=1}^L F(m_i,m_{i+1}) = 0, \label{fmi} 
\eq
with $m_{L+1} \equiv m_1$, which holds for any numbers $m_1,\dots,m_L$ with 
$L>2$,
is $F(a,b)=h(a)-h(b)$ . The first step is to take derivatives with respect 
to $m_j, m_{j+1}$, which leads us to
\bq
\frac{\partial^2 F(m_j,m_{j+1})}{\partial m_j \partial m_{j+1}} = 0. 
\label{dfmn}
\eq
Integrating the above equation over $m_j, m_{j+1}$ we obtain that $F(m,n)$ 
which obeys (\ref{dfmn}), can be most generally expressed as
\bq
F(m,n) = h_1(m) + h_2(n), \label{fmn2}
\eq
with some arbitrary functions $h_1(m),h_2(n)$. The form (\ref{fmn2}) is only 
a necessary condition to have Eq.~(\ref{fmi}) fulfilled. To find the 
sufficient condition, we insert (\ref{fmn2}) into (\ref{fmi}) and obtain
\bq
\sum_i \left[h_1(m_i) + h_2(m_i) \right]= 0.
\eq
Varying one of $m_i$'s while keeping other variables fixed we see that it 
must be $h_1(m)=-h_2(m)$. Defining now $h(m)\equiv h_1(m)$ we see that 
$F(m,n)=h(m)-h(n)$ which concludes our proof.

Interestingly, one can also show in a similar way that if 
$F(a_1,a_2,\dots,a_N)$ depends on $N>2$ variables, then the most general 
solution to the equation
\bq
\sum_{i=1}^L F(m_i,m_{i+1},\dots,m_{i+N-1}) =0
\eq
is $F(a_1,a_2,\dots,a_N)=h(a_1,a_2,\dots,a_{N-1})-h(a_2,a_3,\dots,a_N)$, 
provided that we demand the solution to be the same, irrespective of the 
number of sites $L>N$.

\section*{Appendix B}
We shall show that the condition (\ref{miscond})
\bq
u(m,n) = u(m+1,n-1) \frac{u(1,m)u(n,0)}{u(m+1,0)u(1,n-1)} + u(m,0) - u(n,0) 
\label{b0}
\eq
on the factorization of the steady state in the misanthrope process reduces 
in fact to two equations (\ref{c1}) and (\ref{c2}):
\ba u(n,m) = u(m+1,n-1)
  \frac{u(1,m)u(n,0)}{u(m+1,0)u(1,n-1)}, \label{bc1} \\
  u(n,m) - u(m,n) = u(n,0)-u(m,0). \label{bc2}
\ea
We will prove Eqs.~(\ref{bc1}) and (\ref{bc2}) by induction. Let us first 
replace $n$ by $n=m+p$ in Eq.~(\ref{bc2}), which gives
\bq
u(m+p,m)-u(m,m+p) = u(m+p,0) - u(m,0). \label{b1}
\eq
It follows then from Eq.~(\ref{b0}) that
\ba
u(m,m+p) = & u(m+1,m+p-1) \frac{u(1,m)u(m+p,0)}{u(m+1,0)u(1,m+p-1)} 
\nonumber \\
&+ u(m,0) - u(m+p,0).
\ea
Using Eq.~(\ref{b1}) and changing variables $m\to m-1,p\to p+2$ we obtain
\bq
u(m,m+p) = u(m+p+1,m-1) \frac{u(1,m+p)u(m,0)}{u(m+p+1,0)u(1,m-1)}. 
\label{b2}
\eq
This equation stems from (yet-to-be proved) Eqs.~(\ref{b1}) and known 
Eq.~(\ref{b0}). It is therefore enough to prove only Eq.~(\ref{b1}) by 
induction. Equation (\ref{b1}) is clearly fulfilled for $p=0$. For $p=1$, 
which corresponds to $n=m+1$, we have from Eq.~(\ref{b0}) that
\ba
u(m,m+1) &=& u(m+1,m) \frac{u(1,m)u(m+1,0)}{u(m+1,0)u(1,m)} + u(m,0) - 
u(m+1,0) \nonumber \\
&=& u(m+1,m) + u(m,0) -  u(m+1,0),
\ea
which is equivalent to Eq.~(\ref{b1}) with $p=1$. We thus know that 
Eq.~(\ref{b1}) holds for $p=0,1$. To prove it for arbitrary $p$, we again 
use Eq.~(\ref{b0}) with $m\to m-1,n\to m+p+1$:
\ba
u(m-1,m+p+1) &=& u(m,m+p) \frac{u(1,m-1)u(m+p+1,0)}{u(m,0)u(1,m+p)} 
\nonumber \\
&+& u(m-1,0) - u(m+p+1,0),
\ea
and by inserting $u(m,m+p)$ from Eq.~(\ref{b2}) we obtain
\bq
u(m-1,m+p+1) = u(m+p+1,m-1) + u(m-1,0) - u(m+p+1,0),
\eq
which is equivalent to Eq.~(\ref{b1}) but with $m,p$ replaced by $m-1$ and 
$p+2$, respectively. We have thus proved that equations (\ref{b1}) and 
(\ref{b2}) are fulfilled for any $p$ which completes our proof.

\section*{Appendix C}
We will prove that the Misanthrope process cannot be extended to a 
pair-factorized steady state by relaxing the constraint (\ref{miscond}) (or, 
equivalently, Eqs.~(\ref{c1},\ref{c2})) on the hop rate. We start by 
supposing that this is not the case and one can find $g(m,n)$ such that the 
equation for the probability current (\ref{pbalance}) is fulfilled by
\bq
P(m_1,\dots,m_L) = \prod_{i=1}^L g(m_i,m_{i+1}).
\eq
Following the same lines as in Sec. \ref{mis} and using results of Appendix 
A for $N=3$, we arrive at a relation between $u(m,n), g(m,n)$ and an 
auxiliary function $h(m,n,k)$:
\ba
u(l,m) - u(l+1,m-1)\frac{g(k,l+1)g(l+1,m-1)g(m-1,n)}{g(k,l)g(l,m)g(m,n)} 
\nonumber \\
= h(k,l,m)-h(l,m,n). \label{hmn2}
\ea
Our first task is to express $h(k,l,m)$ through the hop rate. We will do 
this in three steps. First, by assuming $m=0$ we have from Eq.~(\ref{hmn2}) 
that
\bq
u(l,0) = h(k,l,0) - h(l,0,n). \label{st1}
\eq
The left-hand site depends on $l$ only and hence it must be that $h(k,l,0$) 
and $h(l,0,n)$ depend only on their second and first arguments, 
respectively. Assuming $l=0$ in Eq.~(\ref{hmn2}) we obtain that
\bq
0 = u(1,m-1) \frac{g(k,1)g(1,m-1)g(m-1,n)}{g(k,0)g(0,m)g(m,n)} + h(k,0,m) - 
h(0,m,n). \label{st12}
\eq
However, we have just shown that $h(k,0,m)$ is a function of $k$ only. The 
above equation thus depends on $k$ only through $h(k,0,m)$ and 
$g(k,1)/g(k,0)$, the last term being multiplied by the function of $m,n$. 
The equation has to be valid for any $m,n,k$, and hence it must be 
$h(k,0,m)\equiv H=\rm const$ and $g(k,1)/g(k,0)\equiv G=\rm const$.
We can now calculate the ratio $g(m-1,n)/g(m,n)$ from Eq.~(\ref{st12}):
\bq
\frac{g(m-1,n)}{g(m,n)} = G \frac{\left[ h(0,m,n) - 
H\right]g(0,m)}{g(1,m-1)u(1,m-1)}. \label{st2}
\eq
Assuming now $n=0$ and inserting Eq.~(\ref{st2}) into Eq.~(\ref{hmn2}) we 
obtain
\ba
u(l,m) - u(l+1,m-1) \frac{g(k,l+1)g(l+1,m-1)}{g(k,l)g(l,m)} \nonumber \\
\times G \frac{h(0,m,0)-H}{u(1,m-1)} \frac{g(0,m)}{g(1,m-1)} \nonumber  \\
= h(k,l,m) - h(l,m,0).
\ea
From Eq.~(\ref{st1}) we obtain that $h(l,m,0)=u(m,0)+H$ which leads to the 
expression for $h(k,l,m)$:
\ba
h(k,l,m) = u(m,0)+H+u(l,m) \nonumber \\
-u(l+1,m-1) 
G\frac{g(k,l+1)g(l+1,m-1)g(0,m)u(m,0)}{g(k,l)g(l,m)g(1,m-1)u(1,m-1)}. 
\label{st3}
\ea
Next, we insert $h(k,l,m)$ from Eq.~(\ref{st3}) into the initial equation 
(\ref{hmn2}) and, after some calculations, we obtain
\ba
\frac{g(k,l+1)}{g(k,l)} 
u(l+1,m-1)\frac{g(l+1,m-1)Gg(0,m)}{g(l,m)u(1,m-1)g(1,m-1)}\left[u(n,0)+ 
u(m,n)\right. \nonumber \\
\left. - 
u(m+1,n-1)\frac{g(0,m+1)g(m+1,n-1)g(0,n)Gu(n,0)}{g(0,m)g(m,n)g(1,n-1)u(1,n-1)} 
 - u(m,0) \right] \nonumber \\
= \left[ u(n,0)+u(m,n) - u(m,0)\right. \nonumber \\
\left. - 
u(m+1,n-1)\frac{g(l,m+1)g(m+1,n-1)g(0,n)Gu(n,0)}{g(l,m)g(m,n)g(1,n-1)u(1,n-1)} 
\right].
\ea
The only dependence on $k$ in the above equation is through the ratio 
$g(k,l+1)/g(k,l)$. To be valid for all $k$, this ratio has to be 
$k$-independent or the expressions in square brackets must vanish. In the 
first case, the only possibility is that $g(k,l)$ factorizes into a product 
of two functions of separately $k$ and $l$. In the second case, the only 
dependence on $l$ in the right-hand side square bracket is through 
$g(l,m+1)/g(l,m)$. Using the same reasoning we again obtain the 
factorization of $g(l,m)$. Inserting a factorized form, e.g., 
$g(m,n)=\sqrt{f(m)f(n)}$ into the above equation, one arrives at the 
constraint (\ref{miscond}) for the fully factorized misanthrope process. We 
thus see that the extension of the misanthrope process to pair-factorized 
steady states by relaxing the constraint (\ref{miscond}) is impossible.

\end{document}